\documentclass[a4paper,12pt]{amsart}
\usepackage{caption}
\usepackage{cite}
\usepackage{fourier}
\usepackage[margin=3cm]{geometry}
\usepackage{graphicx}
\usepackage{harvard}
\usepackage{hyperref}
\usepackage[hang]{subfigure}
\usepackage{float}
  \restylefloat{figure}
\usepackage{mathrsfs}

\usepackage{brunnian}
  \usetikzlibrary{arrows}
  \usetikzlibrary{external}

\newcommand{\Ll}{\mathscr{L}}
\newcommand{\Pp}{\mathcal{P}}
\newcommand{\R}{\mathbb{R}}

\theoremstyle{plain}
\newtheorem*{theorem*}{Theorem}

\theoremstyle{definition}
\newtheorem{definition}[subsection]{Definition}
\newtheorem{example}[subsection]{Example}
\newtheorem*{example*}{Example}
\newtheorem*{acknowledgements}{Acknowledgements}

\makeatletter
\def\ScaleIfNeeded{%
\ifdim\Gin@nat@width>\linewidth
0.65\linewidth
\else
\Gin@nat@width
\fi
}

\begin{document}
\title{New States of Matter Suggested By New Topological Structures}
\author[N.A. Baas]{Nils A.\ Baas$^\ast$}
\dedicatory{Deparment of Mathematical Sciences, NTNU, N-7491
  Trondheim, Norway\\ \mbox{}\\ (Received 25 June 2012; final version
  received 21 August 2012)}
\thanks{$^\ast$Email: baas@math.ntnu.no}

\maketitle

\begin{abstract}
  We extend the well-known Borromean and Brunnian rings to new higher
  order versions.  Then we suggest an extension of the connection
  between Efimov states in cold gases and Borromean and Brunnian rings
  to these new higher order links.  This gives rise to a whole new
  hierarchy of possible states with Efimov states at the bottom.\\

\end{abstract}

\section{Introduction}
In recent years some strange and counterintuitive states of matter
have been observed in cold gases and nuclear systems
\cite{Curtis,DE,DSG,Esry,FJT,Fer,GFJ,Knoop,K,PDH,Ste,Ste2,SDG,YFJ}.
These states are now commonly called Borromean states or Efimov
states.  They were predicted by V.\ Efimov in 1970, \citeasnoun{Efi},
and are stably bound states of three particles but no two of the three
are bound together.  Such three-particle systems are called Efimov
trimers.  There is also strong evidence for more general weakly bound
cluster states with similar properties.  Borromean rings in topology
represent an analogy for this.  They are three rings in three
dimensional space linked together in such a way that no two of the
three are linked. (For the definition of types of links, see Section
\ref{sec:general}.)

\begin{figure}[H]
  \includegraphics[width=\ScaleIfNeeded]{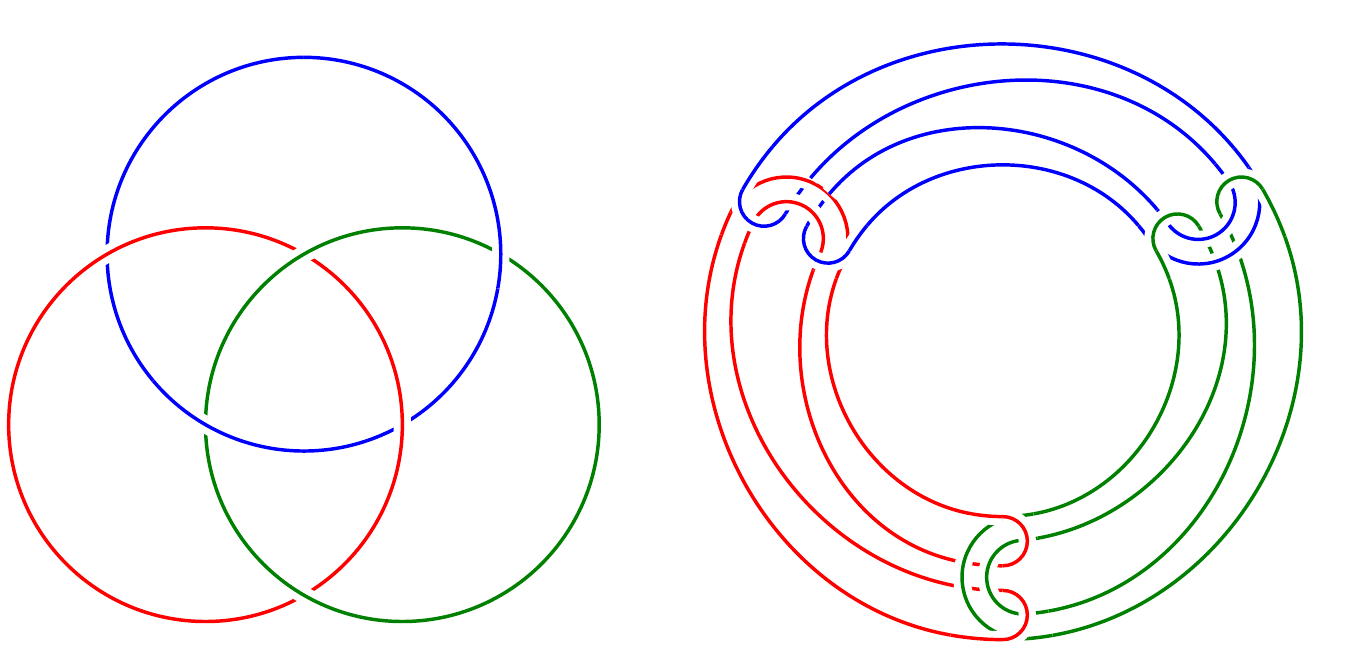}
  \centering

  \caption{Borromean rings to the left and Brunnian rings of type
    $1B(3)$ to the right}
  \label{fig:1}
\end{figure}

(Figures in colour are available at:
\href{http://arxiv.org/pdf/1012.2698.pdf}{http://arxiv.org/pdf/1012.2698.pdf})\\

By analogy we may think of the topological rings as models of the
particles and topological linking as modelling the interactions.  In
other words topological linking corresponds to bound states and
unlinking to unbound states.  Aspects of the basic analogy between
topological entanglement and quantum entanglement have been studied in
\citeasnoun{KL}.

The main purpose of this article is to show how some new higher
order topological linking structures suggest a whole hierarchy of new
states of matter where Efimov states are examples of the lowest
level.

\section{Topological Links}
A topological link is an embedding of circles in three dimensional
space which may be linked together.  A knot is a link with just one
component.

\noindent \textbf{Examples:}\\

\textbf{Trefoil knot:}

\begin{figure}[H]
  \centering
  \begin{tikzpicture}[every path/.style={double=red,knot},every
    node/.style={transform shape,knot crossing,inner
      sep=1.5pt},>=triangle 60]
    \foreach \brk in {0,1,2} {
      \begin{scope}[rotate=\brk * 120]
        \node (k\brk) at (0,-1) {};
      \end{scope}
    }
    \foreach \brk in {0,1,2} {
      \pgfmathparse{int(Mod(\brk - 1,3))}
      \edef\brl{\pgfmathresult}
      \draw (k\brk) .. controls (k\brk.4 north west) and (k\brl.4
        north east) .. (k\brl.center);
      \draw (k\brk.center) .. controls (k\brk.16 south west) and
        (k\brl.16 south east) .. (k\brl);
    }
  \end{tikzpicture}
  \caption{}
  \label{fig:2}
\end{figure}

\textbf{Borromean rings:} see Figure \ref{fig:1}.\\

\textbf{Ring of rings:}

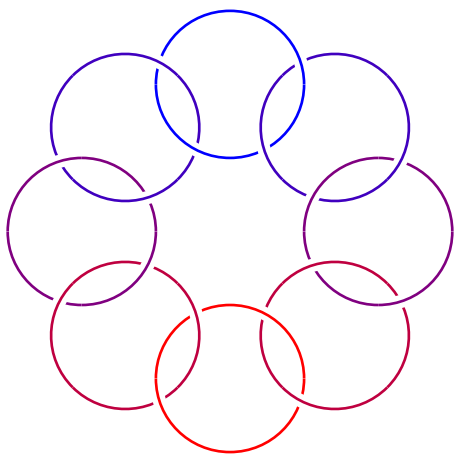
\begin{figure}[H]
  \centering
  \begin{tikzpicture}[scale=0.65]
    \pgfmathsetmacro{\brradius}{1.5}
    \pgfmathsetmacro{\brshift}{2*\brradius }
    \foreach \k in {1,...,8} {
      \pgfmathsetmacro{\brtint}{(\k > 4 ? 8 - \k : \k) * 25}
      \begin{scope}[rotate=\k * 45,yshift=\brshift cm]
        \draw[knot,double=Red!\brtint!Blue] (0,\brradius) arc
          (90:0:\brradius);
        \draw[knot,double=Red!\brtint!Blue] (0,-\brradius) arc
          (-90:-180:\brradius);
        \begin{pgfonlayer}{back}
          \draw[knot,double=Red!\brtint!Blue] (0,\brradius) arc
            (90:180:\brradius);
          \draw[knot,double=Red!\brtint!Blue] (0,-\brradius) arc
            (-90:0:\brradius);
        \end{pgfonlayer}
      \end{scope}
    }
  \end{tikzpicture}
  \caption{Type $1H(8)$}
  \label{fig:3}
\end{figure}

The notion of Borromean rings has been extended to Brunnian rings as
follows.

\begin{definition}
  \label{def:BrunnianRingN}
  $n$ rings in three dimensional space form a Brunnian ring of length
  $n$ if and only if they are linked in such a way that if any
  component is removed the $(n - 1)$ remaining ones are unlinked, or
  in other words, every sublink is trivial.
\end{definition}

Hence Borromean rings are Brunnian of length $3$.  We shall refer to
them all as $B$-structures, see the figure in Example 2 in Section
\ref{sec:newtop}.

The classical Borromean rings have $3$ components.  We may also
introduce Borromean rings of length $n$:

\begin{definition}
  \label{def:BorromeanRingN}
  $n$ rings in three dimensional space form a Borromean ring of length
  $n$ if and only if any two components are unlinked.
\end{definition}

This generalizes to:

\begin{definition}
  \label{def:generalBorromean}
  $n$ rings in three dimensional space form a $B\langle k\rangle$ link
  if and only if any subset of $k$ or fewer rings are unlinked ($1\leq
  k\leq n$).
\end{definition}

Hence:
\begin{align*}
  & B\langle n - 1\rangle \text{ links $=$ Brunnian links of length
    $n$}\\
  & B\langle 2\rangle \text{ links $=$ Borromean links of length
    $n$}\\
  & B\langle 1\rangle \text{ links $=$ $n$ unlinked rings.}\\
\end{align*}

This may be useful in terms of thinking of possible new molecules
and physical states, and extends to higher order as well.

For examples see Figures \ref{fig:37}--\ref{fig:Borrolength4-2} in the
Appendix.  See also \citeasnoun{LM}.

\section{Extended Efimov States}
Both in theory and experiments, extensions of Efimov states to $n$
particle systems, primarily $n = 4$ and $5$ have been considered, see
\cite{DSG,PDH,Ste,Ste2,SDG,YFJ}.  This corresponds metaphorically to
$B$-structures (Brunnian) of length $n$.  These $n$-cluster states are
also sometimes called higher order Efimov states.  We will prefer to
call them just Efimov states of length $n$, or Brunnian states of
length $n$. 
We will just consider the binding properties of the states and ignore
the scaling properties in this paper.

In the following we will introduce topological higher order versions
of Brunnian and Borromean rings.  Since such non-trivial higher order
linkings exist in the topological universe, we think that it is
natural to compare this linking with physical states.  For example a
Brunnian ring of a Brunnian ring of length $3$, would be a linking of
totally $9$ rings (see Figure \ref{fig:7b}).  This raises the
interesting physical question: Does there exist an extended --- second
order --- Efimov state consisting of $9$ particles bound $3$ by $3$
and $3$ clusters (see Figure \ref{fig:19b})?  This is one of the
questions we will discuss in the following sections.

\section{Higher Order Structures}
Often in science one has to consider structures of structures: sets of
sets, vector spaces of vector spaces, forms of forms, links of links,
etc.

In \cite{Baas,NSCS} we introduced a general framework for dealing with
higher order structures, namely what we call Hyperstructures.

In the present paper we will present some geometrical and topological
hyperstructures, namely what we get when we consider links of links of
links $\ldots$\\

\section{New Higher Order Topological Structures}
\label{sec:newtop}
All links considered are in three dimensional space.

Before giving the general construction let us give some examples to
indicate the idea.\\

\noindent \textbf{Example 1:} From rings we may form a ring of rings
based on the Hopf link:

\begin{figure}[H]
  \centering
  \subfigure[Hopf link]{
      \begin{tikzpicture}[scale=0.825]
        \draw[knot,double=ring1] (0,0) arc(180:0:1.5);
        \draw[knot,double=ring2] (3,0) circle (1.5);
        \draw[knot,double=ring1] (0,0) arc(-180:0:1.5);
      \end{tikzpicture}
  } \qquad
  \subfigure[Hopf-ring = ring of rings]{
    \begin{tikzpicture}[scale=0.725]
      \pgfmathsetmacro{\brradius}{.7}
      \pgfmathsetmacro{\brshift}{2*\brradius }
      \foreach \k in {1,...,8} {
        \pgfmathsetmacro{\brtint}{(\k > 4 ? 8 - \k : \k) * 25}
        \begin{scope}[rotate=\k * 45,yshift=\brshift cm]
          \draw[knot,double=Red!\brtint!Green] (0,\brradius) arc
            (90:0:\brradius);
          \draw[knot,double=Red!\brtint!Green] (0,-\brradius) arc
            (-90:-180:\brradius);
          \begin{pgfonlayer}{back}
            \draw[knot,double=Red!\brtint!Green] (0,\brradius) arc
              (90:180:\brradius);
            \draw[knot,double=Red!\brtint!Green] (0,-\brradius) arc
              (-90:0:\brradius);
          \end{pgfonlayer}
        \end{scope}
      }
    \end{tikzpicture} 
  }
  \caption{}
  \label{fig:4}
\end{figure}

Let us call it a Hopf-ring.  From Hopf-rings we may in
the same way form a Hopf-ring of Hopf-rings, etc.\\

\begin{figure}[H]
  \centering
  \subfigure[ring of rings (type $1H(9)$)]{
      \begin{tikzpicture}
        \setbrstep{.1}
        \hopfring{2}{9}
      \end{tikzpicture}
  } \qquad \qquad
  \subfigure[ring of ring of rings (type $2H(4,10)$)]{
    \begin{tikzpicture}[knot/.style={thin knot}]
      \setbrstep{.05}
      \pgfmathsetmacro{\last}{10}
      \pgfmathsetmacro{\angle}{360/10}
      \foreach \l in {1,...,\last} {
        \begin{scope}[rotate=\angle * \l,yshift=-2.5cm]
          \hopfring{1.2}{4}
        \end{scope}
      }
    \end{tikzpicture}
  }
  \caption{}
  \label{fig:5}
\end{figure}

\noindent \textbf{Example 2:} Let us look at the following model of
Brunnian rings, or $B$-rings, of length $n = 3$ and $4$:

\begin{figure}[H]
  \centering
  \includegraphics[width=8.6cm]{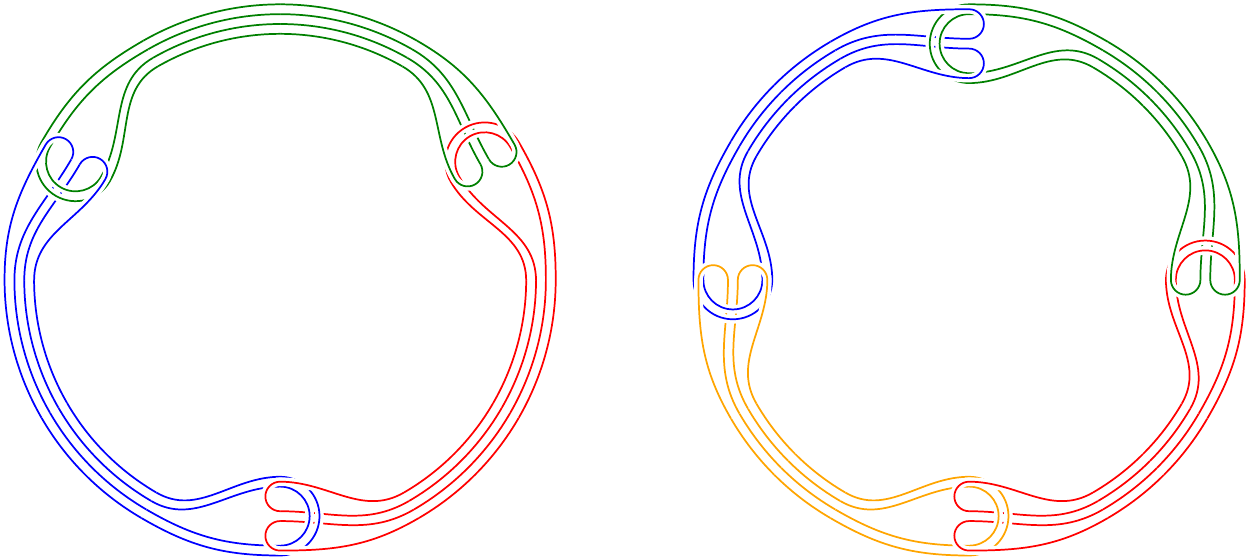}
  \caption{Type $1B(3)$ and $1B(4)$}
  \label{fig:ex2}
\end{figure}

Clearly they are topologically ``ring-like'' in the sense that they
have been formed by chains being made into loops.

But then from these $B$-rings we may form $B$-rings of $B$-rings,
which we shall call $2B$-rings.  This process clearly iterates to
$\mathit{nB}$-rings, and they are truly $n$-th order $B$-rings.

\begin{figure}[H]
  \centering
  \subfigure[$B$-rings (type $1B(3)$)]{
    \includegraphics[scale=0.13]{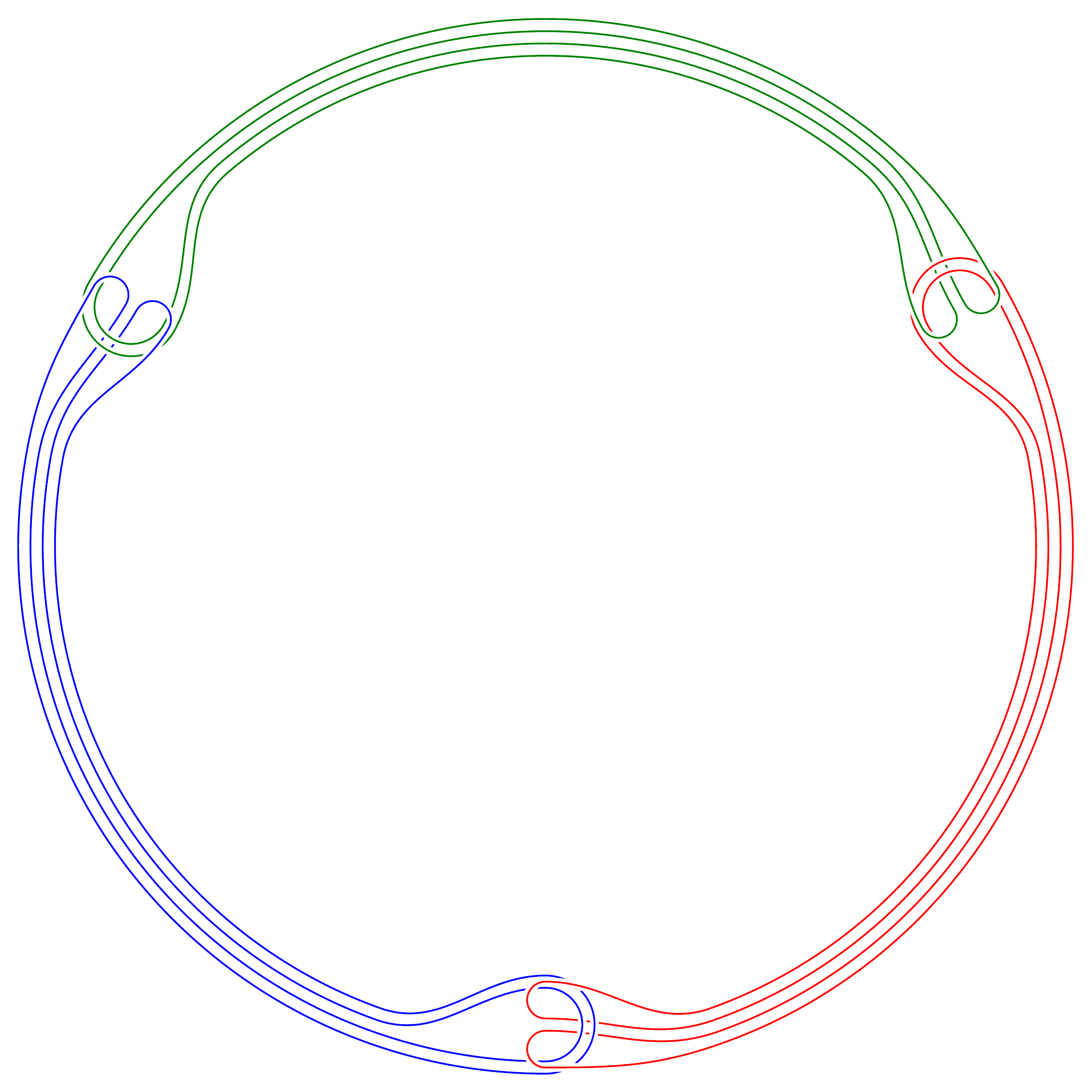}
  } \quad \raisebox{1.25cm}[0pt][0pt]{
    \begin{tikzpicture}[scale=0.65]
      \useasboundingbox (-.1,-.5) rectangle (1.6,.5);
      \draw[ultra thick,->] (0,0) -- (1.5,0);
    \end{tikzpicture}
  } \quad
  \subfigure[$2B$-rings (type $2B(3,3)$)]{
    \includegraphics[scale=0.13]{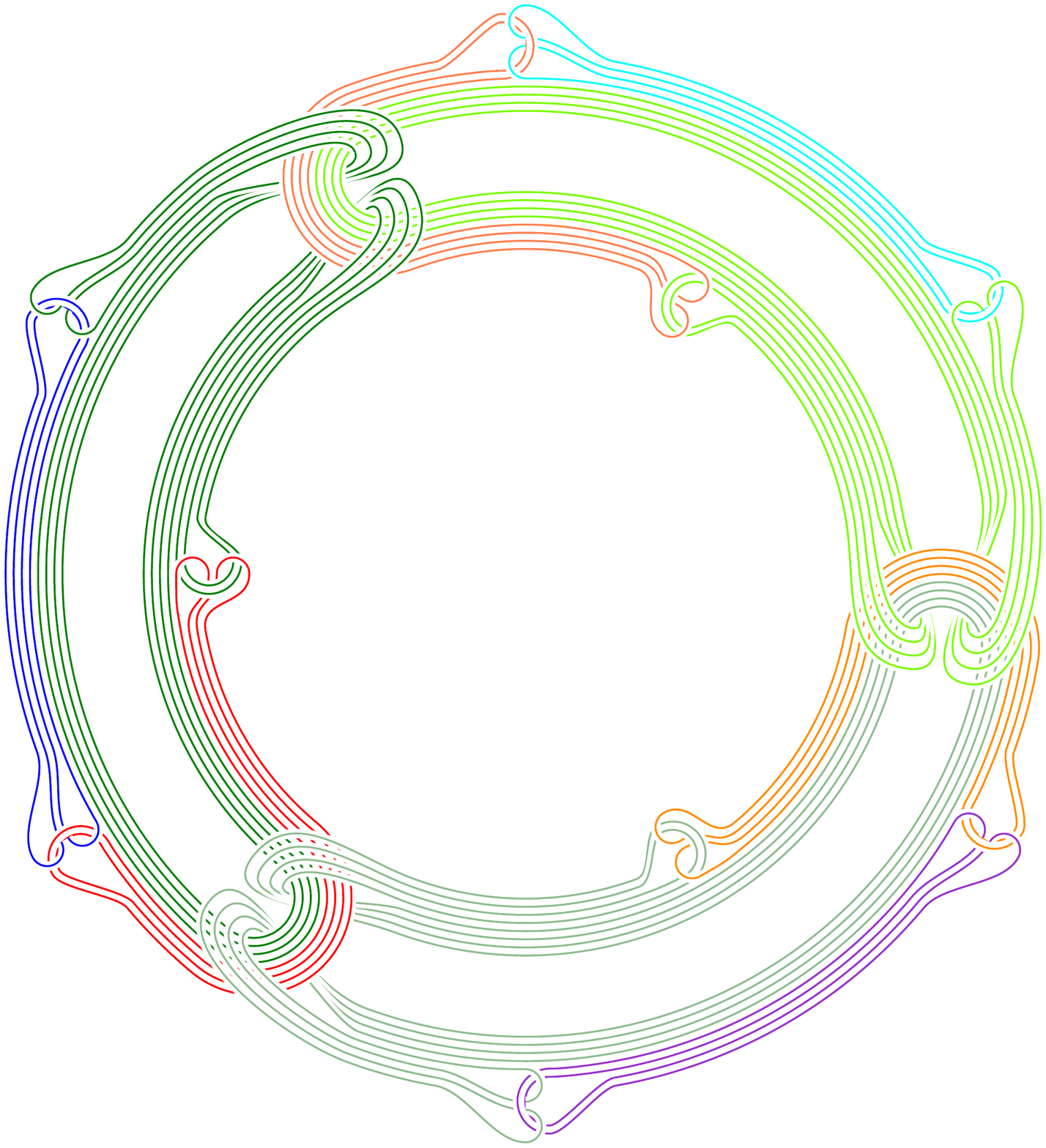}
    \label{fig:7b}
  }
  \caption{}
  \label{fig:Bto2B}
\end{figure}

\begin{figure}[H]
  \centering
  \subfigure[$B$-rings (type $1B(3)$)]{
    \includegraphics[scale=0.13]{ringer/brunnian020_3-crop}
  } \quad \raisebox{1.25cm}[0pt][0pt]{
    \begin{tikzpicture}[scale=0.65]
      \useasboundingbox (-.1,-.5) rectangle (2.1,.5);
      \draw[ultra thick,->] (0,0) -- (2,0);
    \end{tikzpicture}
  } \quad
  \subfigure[$B$-rings of $2$-rings (type $2\mathit{BH}(3,3)$)]{
    \includegraphics[scale=0.0775]{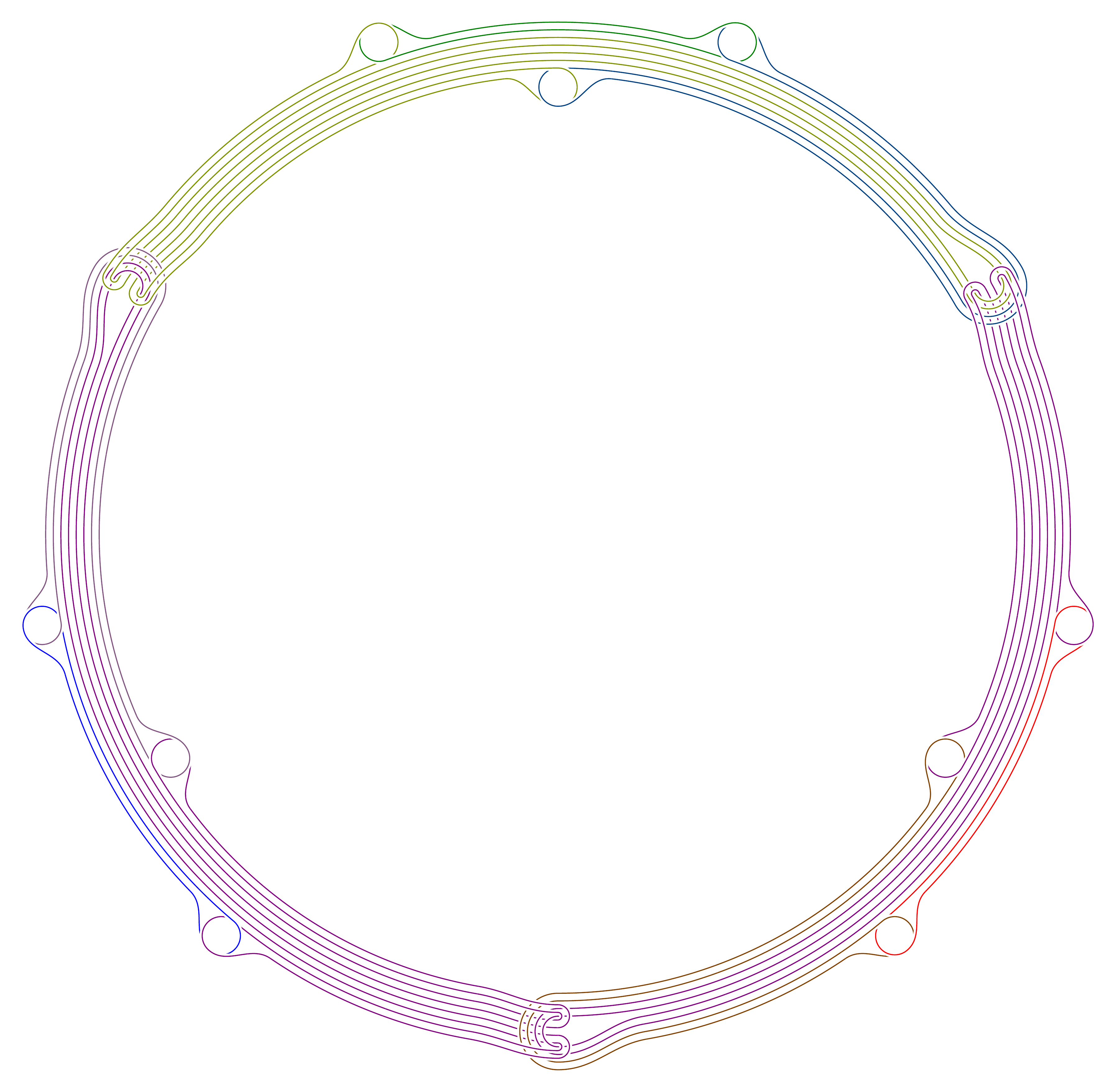}
  }
  \caption{}
  \label{fig:BtoBof2R}
\end{figure}

\begin{figure}[H]
  \centering
  \subfigure[$B$-rings (type $1B(3)$)]{
    \includegraphics[scale=0.14]{ringer/brunnian020_3-crop}
  } \quad \raisebox{1.25cm}[0pt][0pt]{
    \begin{tikzpicture}[scale=0.5]
      \useasboundingbox (-.1,-.5) rectangle (2.1,.5);
      \draw[ultra thick,->] (0,0) -- (2,0);
    \end{tikzpicture}
  } \quad
  \subfigure[rings of $B$-rings (type $2\mathit{HB}(3,3)$)]{
    \includegraphics[scale=0.0775]{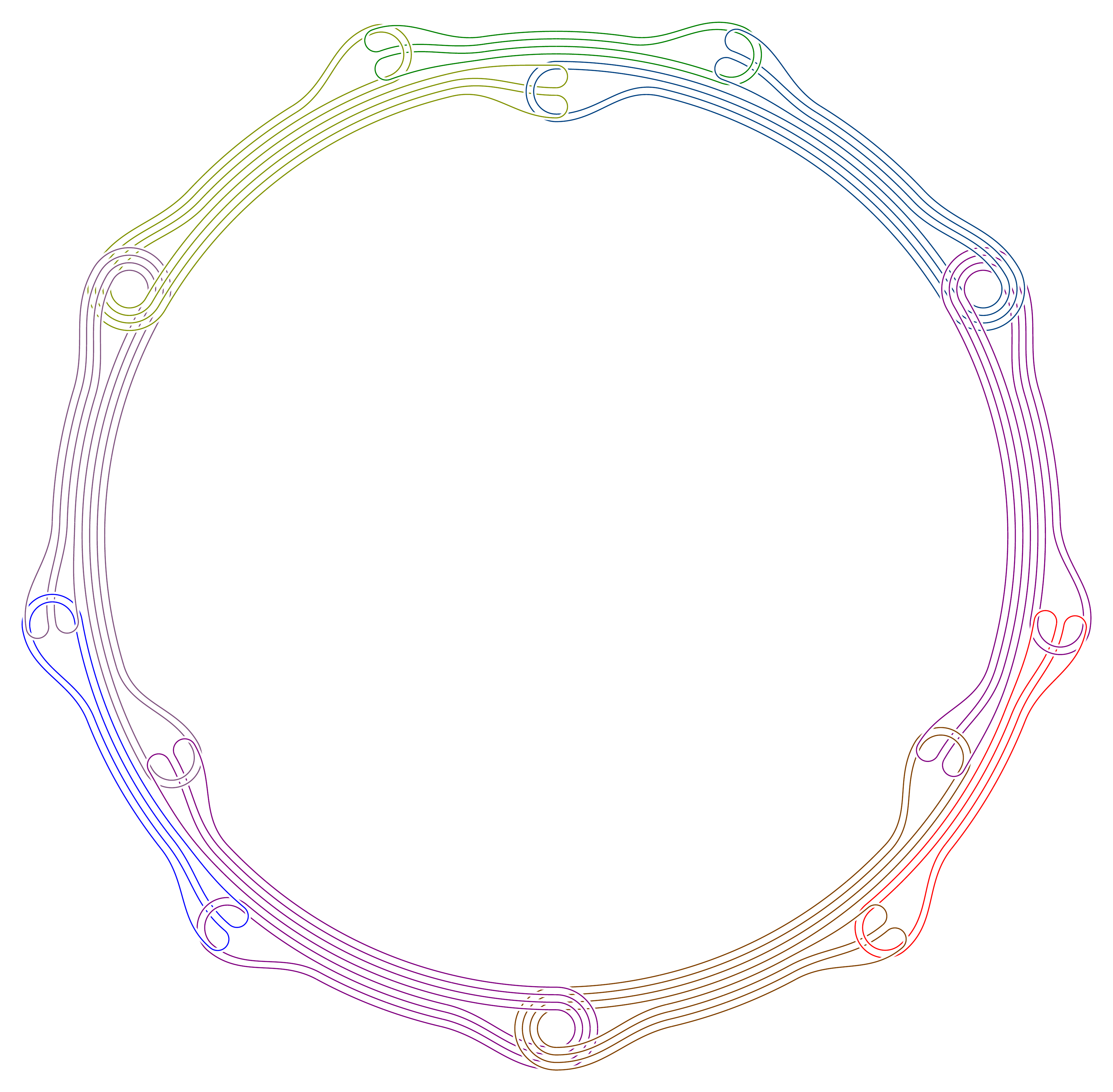}
  }
  \caption{}
  \label{fig:BtoRofBR}
\end{figure}

Just from rings and $B$-rings there are many interesting combinations
of second order structures to be formed:
\begin{align*}
  \text{rings} &\longrightarrow 2\text{-rings} && \text{(see
    Figure \ref{fig:4})}\\
  B\text{-rings} &\longrightarrow 2B\text{-rings} && \text{(see Figure
    \ref{fig:Bto2B})}\\
  B\text{-rings} &\longrightarrow B \text{-rings of 2-rings} &&
  \text{(see Figure \ref{fig:BtoBof2R})}\\
  B\text{-rings} &\longrightarrow \text{rings of $B$-rings} &&
  \text{(see Figure \ref{fig:BtoRofBR})}.
\end{align*}

The geometric and topological properties of these new links are
measured by their complements in three dimensional space:
\begin{equation*}
  \mathit{nK} = \R^3 - (\mathit{nB}).
\end{equation*}
The topology of this is surprisingly complicated even for $n = 2$ as
pointed out in \citeasnoun{NSCS}.  One needs a form of higher order
cohomology operations (Massey products) in order to study them for $n
= 2$.  For $n > 2$ and other links they have not been studied at all.
Higher order Massey products will be needed here.

Higher order links in this new sense have not even been defined nor
studied systematically in the literature prior to \citeasnoun{NSCS},
see also \citeasnoun{Dugo}.  Let us now explain the general link
construction that follows from the hyperstructure idea.\\

\section{The General Idea}
\label{sec:general}
Let $\Ll_1$ be a family of links in $\R^3$, for example rings or
Brunnian rings or both.

Pick a finite number of links from $\Ll_1$ and link or arrange them
together in a chain.

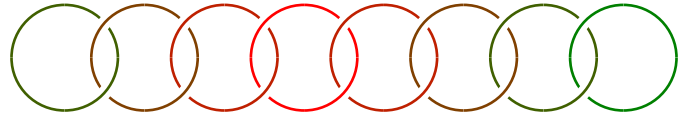
\begin{figure}[H]
  \centering
  \begin{tikzpicture}
    \pgfmathsetmacro{\brradius}{.7}
    \pgfmathsetmacro{\brshift}{1.5*\brradius }
    \foreach \brk in {1,...,8} {
      \pgfmathsetmacro{\brtint}{(\brk > 4 ? 8 - \brk : \brk) * 25}
      \begin{scope}[xshift=\brk * \brshift cm]
        \draw[knot,double=Red!\brtint!Green] (0,0) arc
          (180:90:\brradius);
        \draw[knot,double=Red!\brtint!Green] (\brradius,-\brradius)
          arc (-90:0:\brradius);
        \begin{pgfonlayer}{back}
          \draw[knot,double=Red!\brtint!Green] (0,0) arc
            (-180:-90:\brradius);
          \draw[knot,double=Red!\brtint!Green] (\brradius,\brradius)
            arc (90:0:\brradius);
        \end{pgfonlayer}
      \end{scope}
    }
  \end{tikzpicture}
  \caption{A Hopf chain of Hopf links}
  \label{fig:Hopf_ch}
\end{figure}

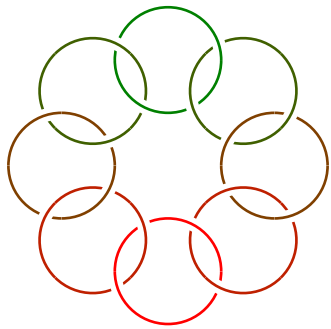
\begin{figure}[H]
  \centering
  \begin{tikzpicture}
    \pgfmathsetmacro{\brradius}{.7}
    \pgfmathsetmacro{\brshift}{2*\brradius }
    \foreach \k in {1,...,8} {
      \pgfmathsetmacro{\brtint}{(\k > 4 ? 8 - \k : \k) * 25}
      \begin{scope}[rotate=\k * 45,yshift=\brshift cm]
        \draw[knot,double=Red!\brtint!Green] (0,\brradius) arc
          (90:0:\brradius);
        \draw[knot,double=Red!\brtint!Green] (0,-\brradius) arc
          (-90:-180:\brradius);
        \begin{pgfonlayer}{back}
          \draw[knot,double=Red!\brtint!Green] (0,\brradius) arc
            (90:180:\brradius);
          \draw[knot,double=Red!\brtint!Green] (0,-\brradius) arc
            (-90:0:\brradius);
        \end{pgfonlayer}
      \end{scope}
    }
  \end{tikzpicture}
  \caption{A Hopf ring formed from a Hopf chain}
  \label{fig:Hopf_ring}
\end{figure}

Use the general deformation principle of rings (O) into U-shaped
figures as follows:

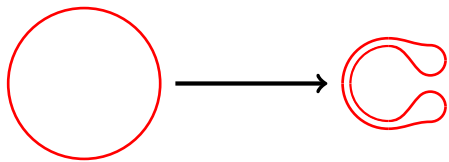
\begin{figure}[H]
  \centering
  \begin{tikzpicture}
    \colorlet{chain}{ring1}
    \draw[knot,double=chain] (-4,0) circle (1);
    \draw[ultra thick,->] (-2.8,0) -- (-.8,0);
    \flatbrunnianlink{1}
  \end{tikzpicture}
  \caption{A Brunnian deformation}
  \label{fig:Brunnian_def}
\end{figure}

\begin{figure}[H]
  \centering
  \begin{tikzpicture}
    \pgfmathsetmacro{\brscale}{1}
    \foreach \brk in {1,...,8} {
      \pgfmathsetmacro{\brtint}{(\brk > 4 ? 8 - \brk : \brk) * 25}
      \pgfmathsetmacro{\brl}{\brscale * \brk}
      \begin{scope}[xshift=\brl cm]
        \colorlet{chain}{Red!\brtint!Green}
        \flatbrunnianlink{\brscale}
      \end{scope}
    }
  \end{tikzpicture}
  \caption{A Brunnian chain}
  \label{fig:Brunnian_chain}
\end{figure}
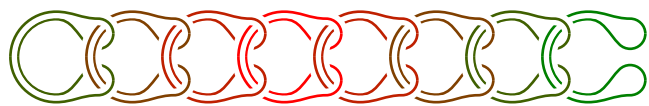

\begin{figure}[H]
  \centering
  \begin{tikzpicture}
    \pgfmathsetmacro{\brscale}{1.8}
    \foreach \brk in {1,...,8} {
      \pgfmathsetmacro{\brtint}{(\brk > 4 ? 8 - \brk : \brk) * 25}
      \begin{scope}[rotate=\brk * 45]
        \colorlet{chain}{Red!\brtint!Green}
        \brunnianlink{\brscale}{45}
      \end{scope}
    }
  \end{tikzpicture}
  \caption{A Brunnian ring}
  \label{fig:Brunnian_ring}
\end{figure}
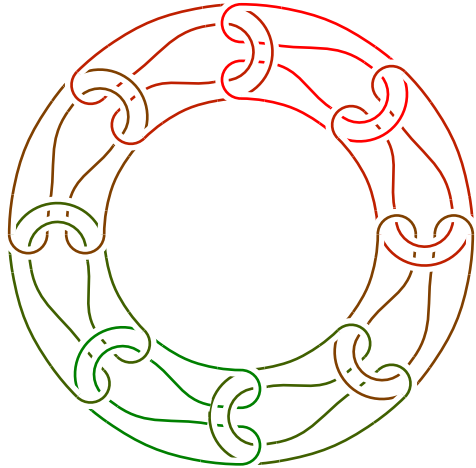



Then form loops (possibly knotted) of such chains in such a way that
they become interlocked.

\begin{figure}[H]
  \centering
  \includegraphics[width=0.8\linewidth]{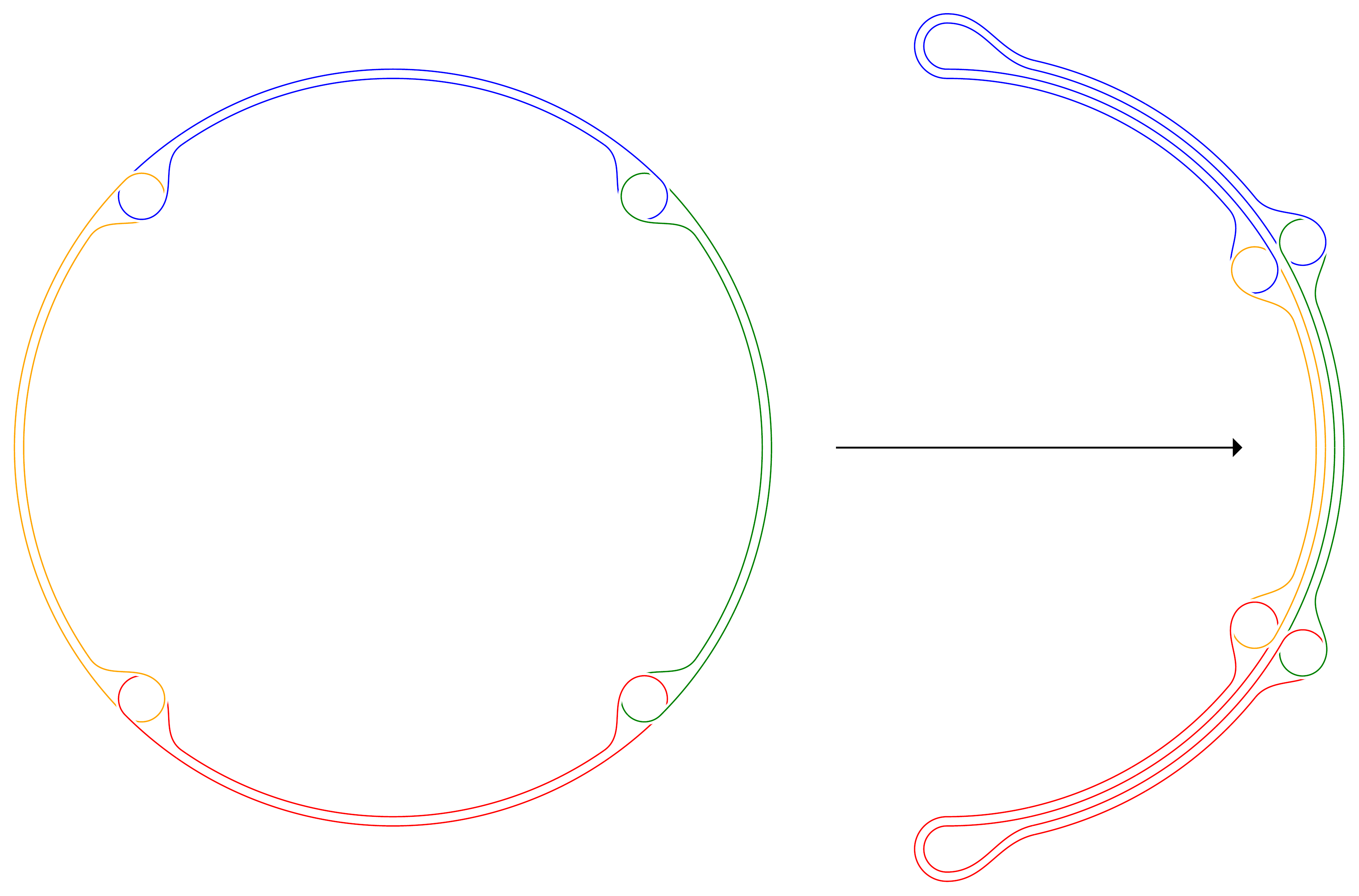}
  \caption{Deformation of a Hopf ring}
  \label{fig:def_RR}
\end{figure}

\begin{figure}[H]
  \centering
  \includegraphics[width=0.8\linewidth]{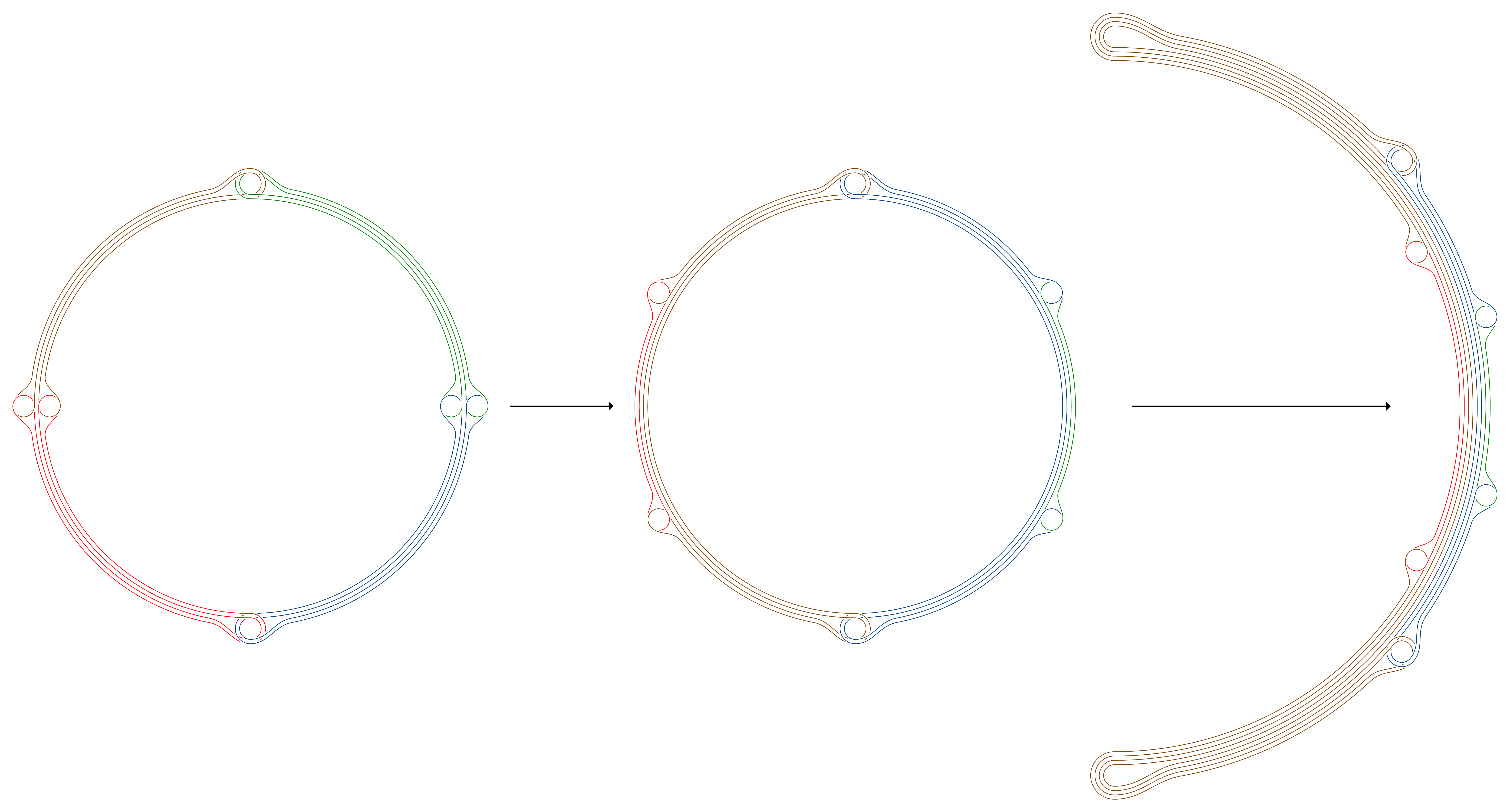}
  \caption{Deformation of a Hopf ring of Hopf rings}
  \label{fig:def_RRR}
\end{figure}

\begin{figure}[H]
  \centering
  \includegraphics[width=0.7\linewidth]{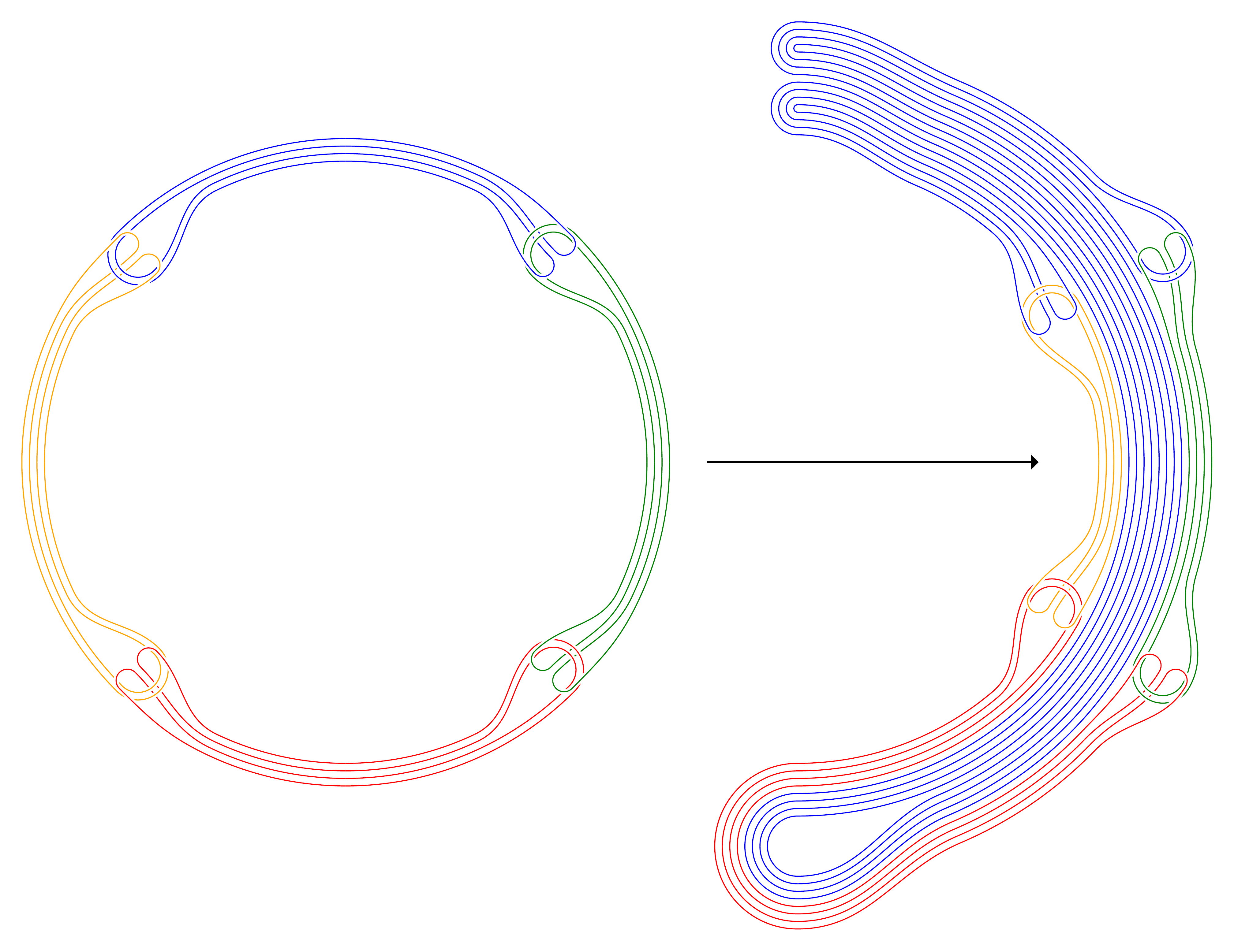}
  \caption{Deformation of Brunnian rings}
  \label{fig:def_BR}
\end{figure}

\begin{figure}[H]
  \centering
  \includegraphics[width=\ScaleIfNeeded]{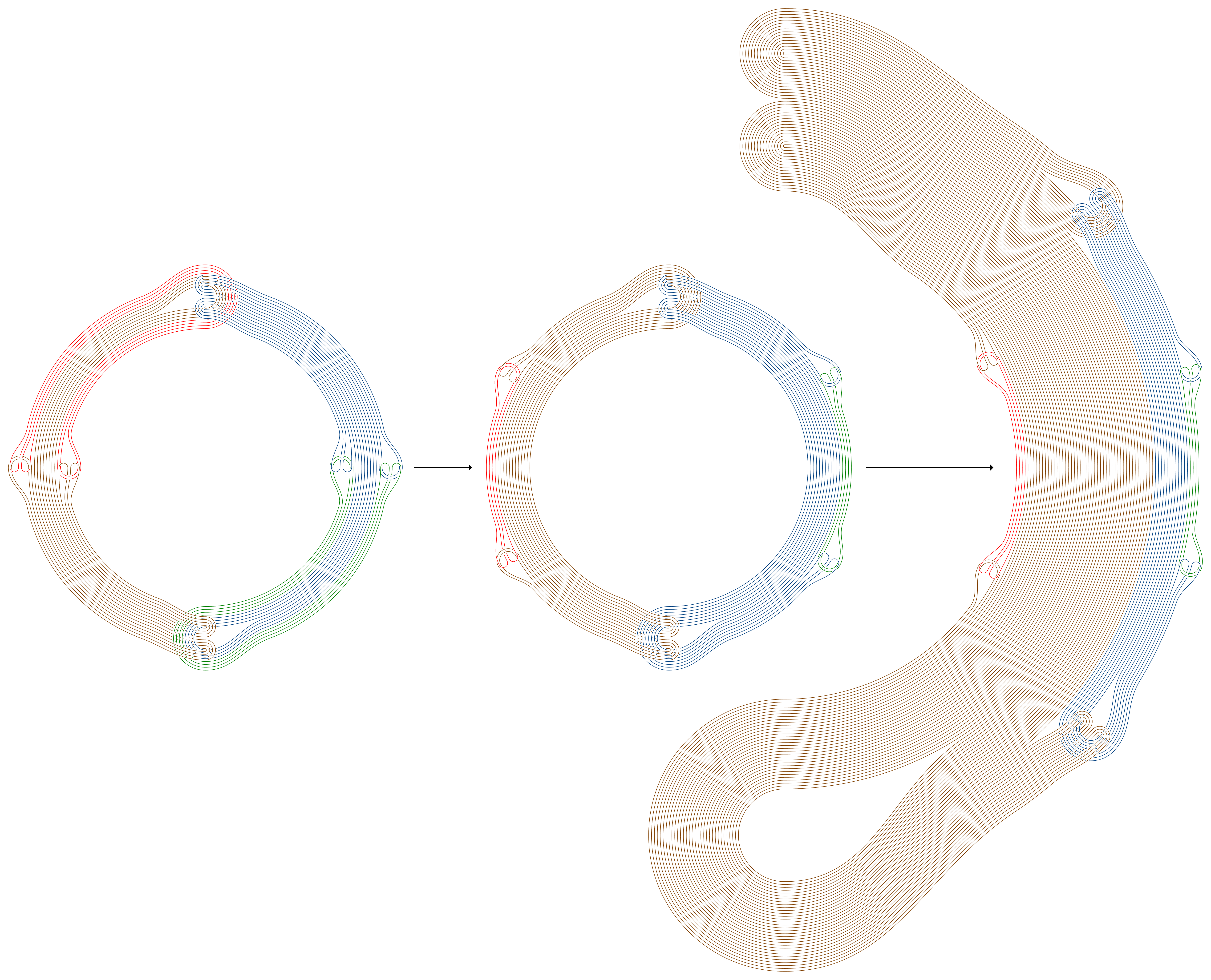}
  \caption{Deformation of a second order Brunnian ring}
  \label{fig:def_Brun_second}
\end{figure}

New links are then formed of these loops (knots) or rings using the
same deformation principle.  We will call these second order links
based on $\Ll_1$ and denote them by
\begin{equation*}
  \Ll_1 \int \Ll_2.
\end{equation*}
Links are formed from the given family $\Ll_1$.  Of course we could
have chosen one family or alphabet of links at each level, but it is
notationally simpler if we keep them all in one family.

Next we apply the same procedure to the family of links $\Ll_1 \int
\Ll_2$, for short $\Ll_2$, but the notation indicate the dependence on
$\Ll_1$, form chains, loops (knots) and new links.  We denote the new
family coming from the constructions
\begin{equation*}
  \Ll_1 \int \Ll_2 \int \Ll_3.
\end{equation*}
This process is then iterated to form
\begin{equation*}
  \Ll_1 \int \Ll_2 \int \ldots \int \Ll_n,
\end{equation*}
abbreviated $\Ll_n$ when no confusion is likely to arise.

\setcounter{subsection}{2}

\begin{definition}
  \label{def:nlink}
  An $n$-th order link is an element of some family $\Ll_n$.
\end{definition}

Clearly our previous examples of $n$-rings and $\mathit{nB}$-rings are
$n$-th order links.  The following definition makes a distinction
between them.

\begin{definition}
  \label{def:nBlink}
  The elements of a family of $n$-th order links
  \begin{equation*}
    \Ll_1 \int \Ll_2 \int \ldots \int \Ll_n
  \end{equation*}
  are called $(n,i)B$ links, if: removing an $\Ll_i$ link from an
  $\Ll_n$ link gives just a collection of unlinked $\Ll_i$ links.
\end{definition}

Clearly $\mathit{nB}$-rings as we have defined them would be
$(n,n-1)B$ links, but $n$-rings would not be.  Rings (closed loops) in
$\R^3$ will be $1$-links.

Let us conclude with some examples.  If $\Ll_1 = \{\text{unknotted
  rings}\}$ then
\begin{align*}
  \text{ring of rings} &= 2 \text{-ring} \in \Ll_1 \int \Ll_2,\\
  B\text{-rings} &\in \Ll_1 \int \Ll_2
\end{align*}
and
\begin{align*}
  2B \text{-rings} &\in \Ll_1 \int \Ll_2 \int \Ll_3,\\
  \text{ring of $B$-rings} &\in \Ll_1 \int \Ll_2 \int \Ll_3,\\
  \text{$B$-rings of $2$-rings} &\in \Ll_1 \int \Ll_2 \int \Ll_3.
\end{align*}
If $\Ll_1 = \{\text{unknotted rings, $B$-rings}\}$, then $2B$-rings
$\in \Ll_1 \int \Ll_2$.\\

The following notation will be useful later on.\\

Let $\Ll_1 = \{\text{unknotted rings}\}$ and let
\begin{equation*}
  \Ll_1^H (n_1) = \{\text{Hopf-rings of length } n_1\},
\end{equation*}
we call these type $1H(n_1)$ links.  Then $\Ll_1^H \int
\Ll_2^H(n_1,n_2)$ will represent Hopf-rings of Hopf-rings, we call
these type $2H(n_1,n_2)$ links and we proceed to form
$\mathit{kH}(n_1,\ldots,n_k)$ links.

Similarly let
\begin{equation*}
  \Ll_1^B(n_1) = \{\text{Brunnian rings of length } n_1\},
\end{equation*}
we call these type $1B(n_1)$ links.  Then $\Ll_1^B \int
\Ll_2^B(n_1,n_2)$ will represent Brunnian rings of length $n_2$ of
Brunnian rings of length $n_1$, we call these type $2B(n_1,n_2)$ links
and we proceed to form $\mathit{kB}(n_1,\ldots,n_k)$ links.

For illustrations of all these types of links see the Appendix.

We have chosen here to illustrate the idea of forming higher order
links by using Brunnian rings of various lengths.  However, one may
also form higher order Borromean rings by for example taking Brunnian
chains and lock them with rings at both ends (see Figure
\ref{fig:38}).  Such Borromean chains may then be linked to form new
rings out of which one may form second order Borromean rings and then
continue the process to any order and with possibly varying lengths at
each level.  We will not pursue this here.  One main point is just to
illustrate the idea of forming higher order links by using higher
order Brunnian links as examples.

Finally, there is an enormous variety of possibilities of forming new
higher order links using the linking and bending (folding) principle
that we have introduced.  Our main purpose has just been to introduce
and illustrate the general principle, and illustrate a few cases.

For example, the bending procedure in Figure \ref{fig:Brunnian_def}
(and Appendix \ref{app:bending}) may be iterated one more time (or
several), then forming new types of chain and rings for further
iteration, see the Appendix.

The general construction here of forming higher order links is a
special case of the process of forming hyperstructures which
encompasses the formation of for example cobordisms of cobordisms
$\ldots$ --- and higher categories, see \cite{Baas,NSCS}.

In the language of knot theory the process may be described by higher
order cables and satellites, embedding the rings in successive tori.
This will be discussed in a separate paper since the main purpose of
this paper is to suggest new physical states.

\section{New States of Matter}
As pointed out in the introduction a remarkable analogy and
correspondence between Borromean rings and Efimov states has recently
been discovered.  Furthermore this extends to Brunnian rings of length
$4$ and $5$, possibly more, and extended Efimov states of the same
length.  This motivates:

\begin{definition}
  \label{def:bstate}
  A Brunnian state of length $n$ is represented by a system of $n$
  particles bound together, but no subsystem is bound.
  
  \noindent \emph{Notice:} No requirements on scaling properties.
\end{definition}

One main goal with this paper is to point out that there is a whole
new and unstudied world of higher order links of the form
\begin{equation*}
  \Ll_1 \int \Ll_2 \int \ldots \int \Ll_n
\end{equation*}
described in the previous section and whose topology is extremely
sophisticated.

\begin{figure}[H]
  \centering
  \subfigure[$1B(3)$-ring $\sim$ trimers (see Figures \ref{fig:Bto2B}
  and \ref{fig:1b3})]{
    \begin{tikzpicture}
      \draw[very thick] (0,0) circle(2.5cm);
      \draw[very thick, blue] (-0.25,-1.6) circle(0.5cm);
      \draw[very thick, red] (-1.1,1.2) circle(0.5cm);
      \draw[very thick, green] (1.4,0.8) circle(0.5cm);
    \end{tikzpicture}
  } \qquad\qquad
  \subfigure[$2B(3,3)$-ring $\sim$ trimers of trimers (see Figures
  \ref{fig:Bto2B} and \ref{fig:2b33})]{
    \begin{tikzpicture}
      \draw[very thick] (0,0) circle(2.5cm);
    
      \begin{scope}[scale=0.3,xshift=-4.25cm,yshift=2.5cm]
        \draw[very thick] (0,0) circle(2.5cm);
        \draw[very thick, blue] (-0.25,-1.6) circle(0.5cm);
        \draw[very thick, red] (-1.1,1.2) circle(0.5cm);
        \draw[very thick, green] (1.4,0.8) circle(0.5cm);
      \end{scope}
    
      \begin{scope}[scale=0.3,xshift=4.5cm,yshift=1.2cm]
        \draw[very thick] (0,0) circle(2.5cm);
        \draw[very thick, blue] (-0.25,-1.6) circle(0.5cm);
        \draw[very thick, red] (-1.1,1.2) circle(0.5cm);
        \draw[very thick, green] (1.4,0.8) circle(0.5cm);
      \end{scope}
    
      \begin{scope}[scale=0.3,xshift=-0.5cm,yshift=-4cm]
        \draw[very thick] (0,0) circle(2.5cm);
        \draw[very thick, blue] (-0.25,-1.6) circle(0.5cm);
        \draw[very thick, red] (-1.1,1.2) circle(0.5cm);
        \draw[very thick, green] (1.4,0.8) circle(0.5cm);
      \end{scope}
    \end{tikzpicture}
    \label{fig:19b}
  }
  \caption{Trimers and trimers of trimers}
  \label{fig:8}
\end{figure}
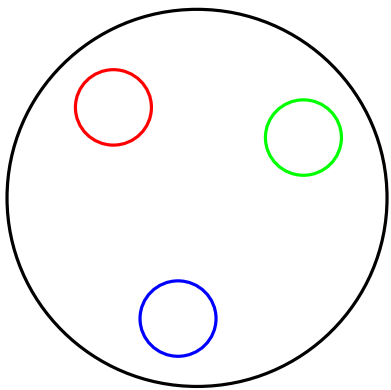
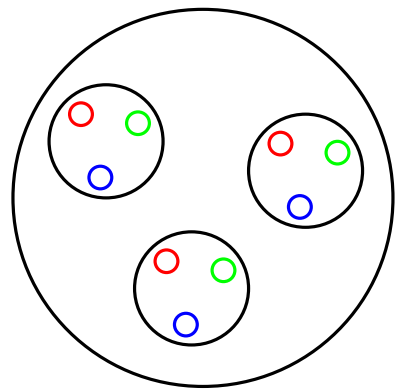

The new idea is then to suggest and predict that there are families of
states of matter (for example in cold gases)
$\Pp_1,\Pp_2,\ldots,\Pp_n$ corresponding to these higher order links,
and organized into the same pattern or hyperstructure.  In other
words:
\begin{equation*}
  \Ll_1 \int \Ll_2 \int \ldots \int \Ll_n \longleftrightarrow \Pp_1
  \int \Pp_2 \int \ldots \int \Pp_n,
\end{equation*}
as in Figure \ref{fig:8}: Trimers and trimers of trimers.

Other possible composite states could for example be:
\begin{itemize}
\item[(i)] A trimer and two singletons:
  \begin{figure}[H]
    \centering
    \begin{tikzpicture}
      \draw[very thick] (0,0) circle(2.5cm);

      \begin{scope}[scale=0.5,xshift=-2cm]
        \draw[very thick] (0,0) circle(2.5cm);
        \draw[very thick, blue] (-0.25,-1.6) circle(0.5cm);
        \draw[very thick, red] (-1.1,1.2) circle(0.5cm);
        \draw[very thick, green] (1.4,0.8) circle(0.5cm);
      \end{scope}

      \draw[very thick,orange] (1.25,1) circle(0.25cm);
      \draw[very thick,brown] (1.25,-1) circle(0.25cm);
    \end{tikzpicture}
    \caption{}
    \label{fig:trimer_singleton}
  \end{figure}
  \noindent which would correspond to the following second order link:
  \begin{figure}[H]
    \centering
    \includegraphics[width=\ScaleIfNeeded]{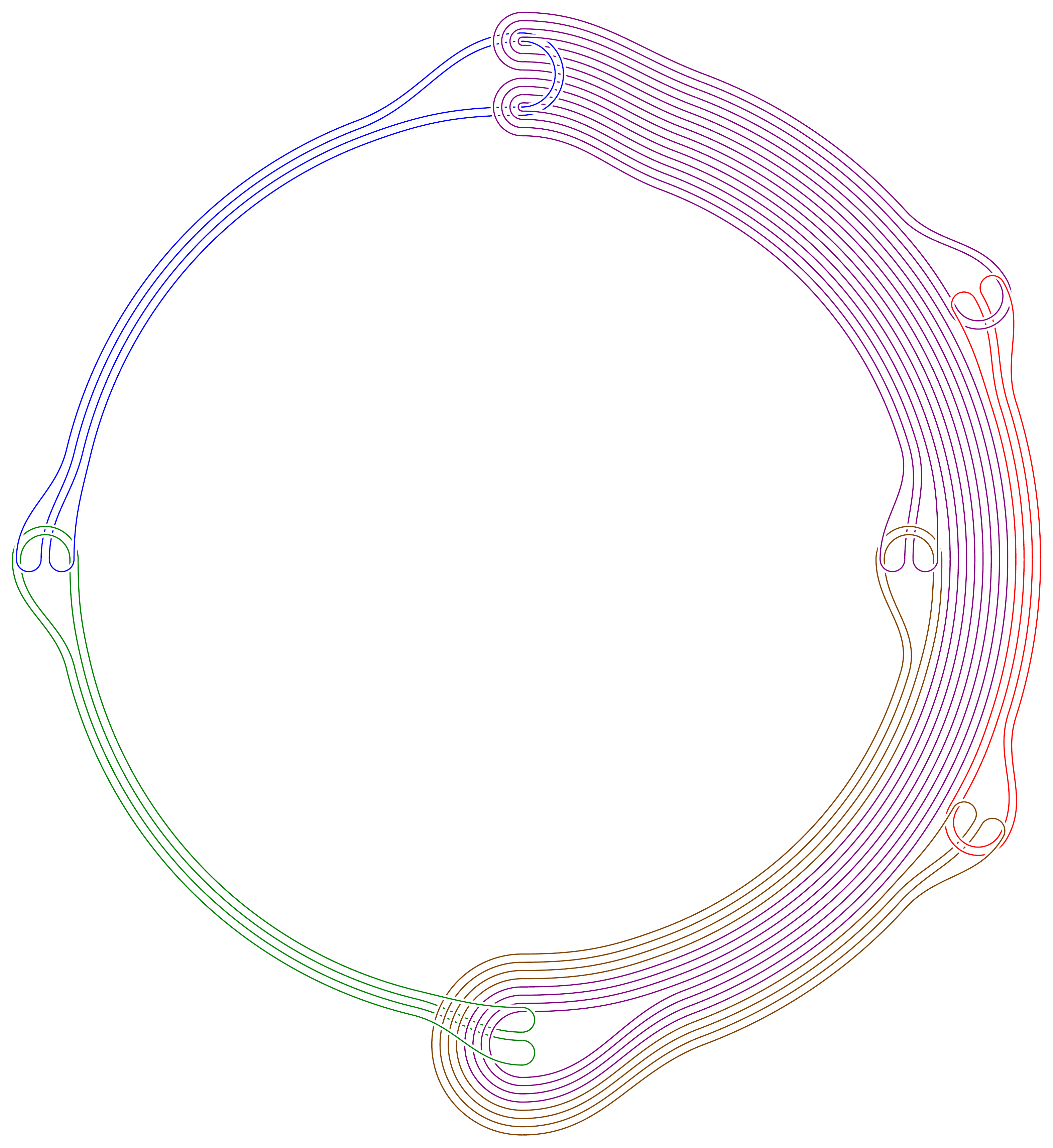}
    \caption{}
    \label{fig:}
  \end{figure}
\item[(ii)] A trimer and a dimer:
  \begin{figure}[H]
    \centering
    \begin{tikzpicture}
      \draw[very thick] (0,0) circle(2.5cm);

      \begin{scope}[scale=0.4,xshift=-2.25cm,yshift=2.5cm]
        \draw[very thick] (0,0) circle(2.5cm);
        \draw[very thick, blue] (-0.25,-1.6) circle(0.5cm);
        \draw[very thick, red] (-1.1,1.2) circle(0.5cm);
        \draw[very thick, green] (1.4,0.8) circle(0.5cm);
      \end{scope}

      \begin{scope}[scale=0.4,xshift=1.95cm,yshift=-1.75cm]
        \draw[very thick] (0,0) circle(2.5cm);
        \draw[very thick,orange] (0,1) circle(0.5cm);
        \draw[very thick,brown] (0,-1) circle(0.5cm);
      \end{scope}
    \end{tikzpicture}
    \caption{}
    \label{fig:trimer_dimer}
  \end{figure}
  \noindent which would correspond to the following second order link:
  \begin{figure}[H]
    \centering
    \includegraphics[width=0.5\linewidth]{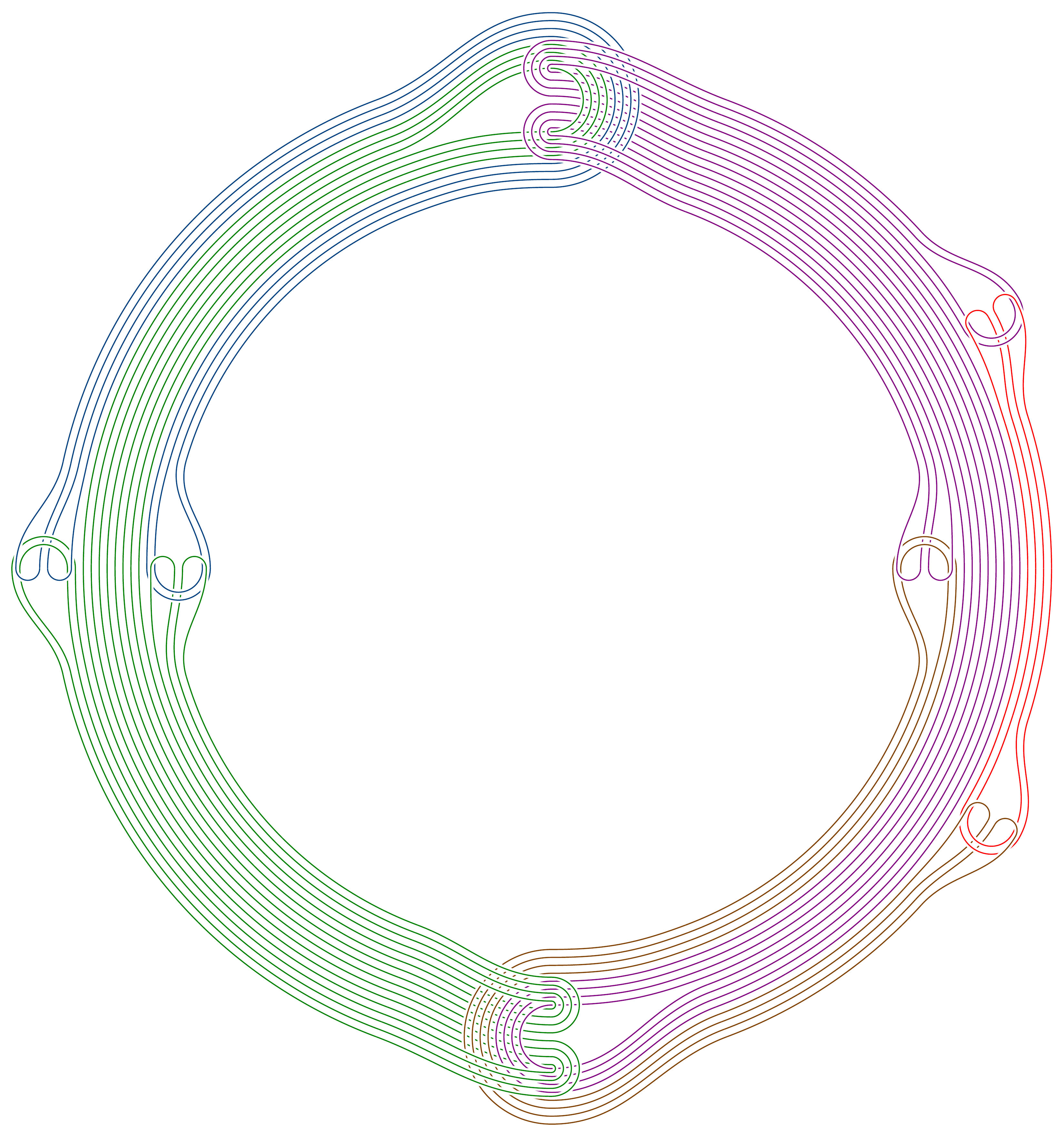}
    \caption{}
    \label{fig:23}
  \end{figure}
\end{itemize}

The links and cluster states in Figures
\ref{fig:trimer_singleton}--\ref{fig:23} have been found in models
discussed in \citeasnoun{Ste2}, and may be considered as intermediate
states between trimers and trimers of trimers.  A trimer of trimers
will clearly require coexistence of two types of resonances: atom-atom
and trimer-trimer.

Based on the general hyperstructure idea in \citeasnoun{Baas} one may
go even further and extend $\Ll_1 \int \Ll_2 \int \ldots \int \Ll_n$
and $\Pp_1 \int \Pp_2 \int \ldots \int \Pp_n$ to include patterns as
in Figure \ref{fig:9}:

\begin{figure}[H]
  \centering
  \includegraphics[width=0.5\linewidth]{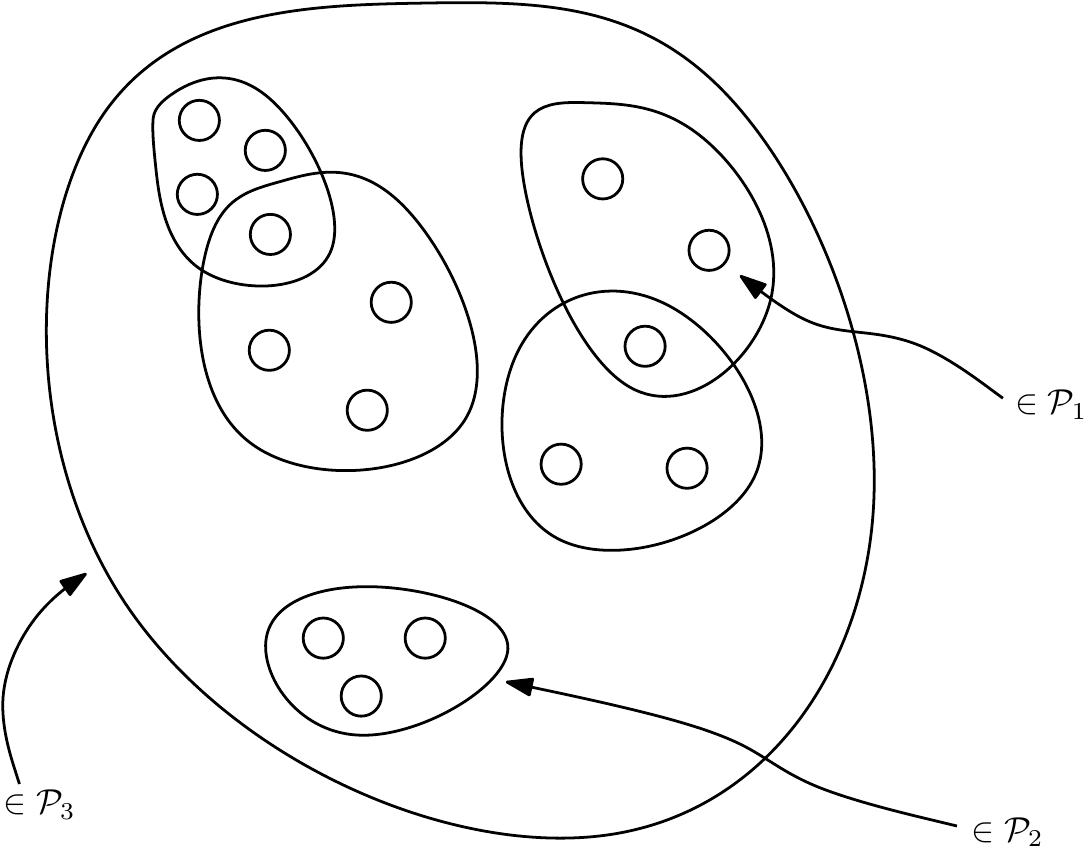}
  \caption{}
  \label{fig:9}
\end{figure}

\noindent and predict their existence.  For further background, see
\cite{Baas,NSCS}.

This means that for example in cold gases one may look for --- by
suitable tuning --- new types of states like: trimers of trimers or
even higher order cluster states as described here.  Furthermore
trimers of tetramers and other combinations may be considered as well.
We know that in cold gases $1B(3)$, $1B(4)$ and $1B(5)$ states have
been found.  What about the existence of $2B(3,3)$, $2B(2,3)$,
$2B(3,2)$, $2B(4,3)$, $3B(3,3,3)$, etc.\ states?  See the Appendix for
the corresponding geometries.

Do there exist weakly bound clusters of particles realizing higher
order Brunnian states?  See \citeasnoun{BFJRVZ} for a discussion of
this question.\\

Let us conclude with our\\

\textbf{Main Prediction:} \emph{To higher order links, in the sense
  that we have defined them, there exist corresponding higher order
  states of matter in physical systems such as cold gases, nuclei and
  other types of many-body systems in general.}\\

Such states of matter may be observed in finely tuned cold gases or
other systems such as Bose--Einstein condensates.  It would also be
interesting to see such states coming directly from a Schr{\"o}dinger
equation with Hamiltonian
\begin{equation*}
  \sum -\frac{\hbar^2}{2m} \nabla_i^2 + V(\text{hyperstructure}),
\end{equation*}
meaning that the potential is determined by the geometric
hyperstructure of links and particles as described here.  One would
then have to take into account levels of cluster-interactions.

It would be very interesting if one could find a connection between
the topology of these higher order links and the quantum entanglement,
see \citeasnoun{KL}.  For a discussion of more general many-body
system interactions, see the addendum or \citeasnoun{manybody}, and a
more extensive discussion in \citeasnoun{structure}.

Let us conclude by mentioning that it is a natural and interesting
question whether the higher order links we have introduced may be
synthesized as molecules.  Trefoil knots and Borromean rings have been
synthesized \cite{DBCS,MSS,See,Stod} by very sophisticated techniques.
To synthesize higher order links is a daunting task, but for example
Borromean rings of Borromean rings and other members of the families
we have defined like $\mathit{kB}(n_1,\ldots,n_k)$ and
$\mathit{kH}(n_1,\ldots,n_k)$ in Section 6 may be possible to
synthesize by using DNA molecules and the techniques developed by N.\
Seeman.  This will be discussed in a separate paper, \citeasnoun{BS}.

\begin{acknowledgements}
  I would like to thank A.\ Stacey for help with the graphics and M.\
  Thaule for technical assistance.  I would also like to thank the
  Institute for Advanced Study, Princeton, USA for the kind
  hospitality during my stay there in the first half of 2010 when
  parts of this work were done.  Furthermore, I would also like to
  thank J.\ D'Incao, V.\ Efimov, R.\ Hulet, D.\ Huse and J.\ von
  Stecher for enlightening discussions and correspondence.
\end{acknowledgements}

\section*{\textbf{Addendum\\[0.5cm] On Many-Body System Interactions}}
We will here discuss possible interactions of families of particles.
By a particle we mean a system or an (extended) object in some space.
The families may be finite, countable or uncountably infinite.

\begin{equation*}
  \Pp = \{P_i\}_{i \in I}
\end{equation*}

What does it mean that the particles interact?

Often we have ``state''-spaces associated to the particles
\begin{equation*}
  P_i \mapsto S_i.
\end{equation*}
An interaction is a rule telling us how the states influence each other
or are related:
\begin{equation*}
  R \subset \prod S_i
\end{equation*}
If this is stable and time independent we may call it a \emph{bond}.\\

\noindent \textbf{Simplicial model:}\\

A \emph{pair} interaction is geometrically represented as:
\raisebox{-0.55cm}[0pt][0pt]{\begin{tikzpicture}
  \draw[thick,font=\scriptsize]
    (0,0) node[below,yshift=-0.10cm]{$P_i$} -- (2,0)
    node[below,yshift=-0.10cm]{$P_j$};
  \draw[thick,fill=black] (0,0) circle(0.10cm);
  \draw[thick,fill=black] (2,0) circle(0.10cm);
\end{tikzpicture}}\\

An $n$-tuple interaction may be represented as an $n$-simplex, but it
is reducible to pair interactions:

\begin{figure}
  \centering
  \begin{tikzpicture}[thick]
    \draw (0,0) -- (3,-1) -- (5,1) -- (2.5,3) -- (0,0);
    \draw (2.5,3) -- (3,-1);
    \draw[dashed] (0,0) -- (5,1);
    \draw (0,0) node[left]{$P_1$};
    \draw (3,-1) node[below]{$P_2$};
    \draw (5,1) node[right]{$P_3$};
    \draw (2.5,3) node[above]{$P_4$};
  \end{tikzpicture}
  \caption{}
  \label{fig:n-simplex}
\end{figure}

Here we represent the systems geometrically by \emph{points} (in some
Euclidean space):

\begin{equation*}
  \text{Particle (system)} \quad \longmapsto \quad \text{point in f.\
    ex.\ $\R^3$}
\end{equation*}

But we may have more sophisticated interactions like:\\

\noindent \textbf{Brunnian or Borromean model:}\\

$n$ particles interact in such a way that no proper subset of them
interacts (Brunnian), or no pair interacts (Borromean), see
Definitions \ref{def:BrunnianRingN}--\ref{def:generalBorromean}.  This
suggests and is best understood by another representation:

\begin{equation*}
  \text{Particle} \quad \longmapsto \quad \text{ring (string) in
    $\R^3$}
\end{equation*}

\begin{figure}[H]
  \centering
  \subfigure[Borromean]{
    \begin{tikzpicture}[every path/.style={knot},scale=0.55]
      \pgfmathsetmacro{\csixty}{cos(60)}
      \pgfmathsetmacro{\ssixty}{sin(60)}
      \pgfmathsetmacro{\brscale}{2}
      \draw[double=ring1] (0,0) circle (\brscale);
      \draw[double=ring2] (\brscale,0) circle (\brscale);
      \begin{pgfonlayer}{back}
        \draw[double=ring3] (\brscale *\csixty,\brscale *\ssixty)
          ++(0,\brscale) arc(90:220:\brscale);
        \draw[double=ring3] (\brscale *\csixty,\brscale *\ssixty)
          ++(0,-\brscale) arc(-90:-10:\brscale);
      \end{pgfonlayer}
      \draw[double=ring3] (\brscale *\csixty,\brscale *\ssixty)
        ++(0,\brscale) arc(90:-10:\brscale);
      \draw[double=ring3] (\brscale *\csixty,\brscale *\ssixty)
        ++(0,-\brscale) arc(-90:-140:\brscale);
    \end{tikzpicture}
  } \qquad \qquad
  \subfigure[Brunnain interactions of length $k = 4$]{
    \begin{tikzpicture}[knot/.style={thin knot},scale=0.55]
      \setbrstep{.1}
      \brunnian{2.5}{4}
    \end{tikzpicture}
  }
  \caption{}
  \label{fig:particle-string}
\end{figure}

In this representation a pair interaction is represented by Hopf
links:

\begin{figure}[H]
  \centering
  \subfigure{
    \begin{tikzpicture}[scale=0.775]
      \draw[knot,double=ring1] (0,0) arc(180:0:1.5);
      \draw[knot,double=ring2] (3,0) circle (1.5);
      \draw[knot,double=ring1] (0,0) arc(-180:0:1.5);
    \end{tikzpicture}
  } \hspace*{-0.05cm} \raisebox{1.2cm}[0pt][0pt]{or} \hspace*{-0.05cm}
  \subfigure{
    \begin{tikzpicture}[scale=0.8]
      \setbrstep{.1}
      \hopfring{2}{2}
    \end{tikzpicture}
  }
  \caption{}
  \label{fig:Hopf-links}
\end{figure}

With this representation we have in the main text introduced a
whole new hierarchy of (possible) higher order interactions
represented by new higher order links of rings.  This has been
extensively developed in the main text.\\

The interesting question is then:\\

\emph{What about other geometric (topological) representations, and
  what kind of new interactions do they suggest --- both of first and
  higher order?}\\

We may proceed as follows.

\begin{align*}
  \text{Particle} \quad \longmapsto \quad \text{Space } & \text{(of
    some kind and}\\
  &\text{f.\ ex.\ embedded in a fixed}\\
  &\text{ambient space $A$)}
\end{align*}

\begin{equation*}
  P_i \longrightarrow T_i = \mathrm{Space}_i \subset A
\end{equation*}

Pictorially this looks like

\begin{figure}[H]
  \centering
  \begin{tikzpicture}[scale=0.5]
    \begin{scope}[rotate=15]
      \foreach \x in {1,5,9}
        \draw[very thick] (\x,0) arc(-50:50:2cm);
      \node[rotate=15] at (3.5,1.75){$\cdots$};
      \node[rotate=15] at (7.5,1.75){$\cdots$};
      \node[below] at (5,-0.5){$T_i$};
      \node[above right] at (12,2){$A$};
    \end{scope}
    
    \begin{scope}[rotate=20]
      \draw[very thick] (5.5,1) ellipse(6cm and 4cm);
    \end{scope}
  \end{tikzpicture}
  \caption{}
  \label{fig:particle-space}
\end{figure}

In the sense of \citeasnoun{Baas} $A$ is a bond of $\{T_i\}$.  This
can be iterated and one may form higher order bonds, ending up with a
\emph{hyperstructure}, defined and described in \citeasnoun{Baas}.

Let us be more specific and consider the situation where the
representing spaces are manifolds embedded in a large ambient manifold
or Euclidean space.

\begin{equation*}
  P_i \longrightarrow M_i \subset B (\subset \R^N)
\end{equation*}

We may have interactions or bonds for example of the following types:

\setcounter{subfigure}{0}

\begin{figure}[H]
  \centering
  \subfigure[Linked]{
    \begin{tikzpicture}[scale=0.8]
      \draw[very thick] (0,0) ellipse(5cm and 3cm);
      \node at (4,2.75){$B$};

      \begin{scope}[xshift=-0.7cm,yshift=-1.25cm,scale=1.1,rotate=15]
        \begin{scope}[scale=0.5,rotate=30,xshift=5.35cm,yshift=-3.35cm]
          \draw[very thick] (-1.1,5.75) arc(90:220:1.5cm and 3.5cm);
          \draw[very thick] (-1.1,5) arc(90:210:1cm and 3cm);
        \end{scope}
      \end{scope}

      \begin{scope}[xshift=1.25cm,scale=1.1]
        \begin{scope}[scale=0.5,rotate=-30,xshift=-1.75cm,yshift=-3.5cm]
          \draw[line width=1ex,white] (0,0) arc(-40:220:1.5cm and 4cm);
          \draw[very thick] (0,0) arc(-40:220:1.5cm and 4cm);
          \draw[line width=1ex,white] (-0.25,0.5) arc(-30:210:1cm and
            3.25cm);
          \draw[very thick] (-0.25,0.5) arc(-30:210:1cm and 3.25cm);
        \end{scope}
        \draw[very thick,rotate=-30] (-1.5,-1.5)
          node[below,yshift=-1cm]{$M_i$} ellipse(1cm and 0.75cm);
      \end{scope}
      
      \begin{scope}[xshift=-0.7cm,yshift=-1.25cm,scale=1.1,rotate=15]
        \begin{scope}[scale=0.5,rotate=30,xshift=5.35cm,yshift=-3.35cm]
          \draw[line width=1ex,white] (0,0) arc(-40:90:1.5cm and 3.5cm);
          \draw[very thick] (0,0) arc(-40:90:1.5cm and 3.5cm);
          \draw[line width=1ex,white] (-0.25,0.5) arc(-30:90:1cm and
            3cm);
          \draw[very thick] (-0.25,0.5) arc(-30:90:1cm and 3cm);
        \end{scope}
        \draw[very thick,rotate=30,xshift=0.15cm] (2,-1.5)
          node[below,yshift=-1cm]{$M_j$} ellipse(1cm and 0.75cm);
      \end{scope}
    \end{tikzpicture}
  } \vspace*{1cm} \subfigure[Connected (by intermediate manifolds)]{
    \begin{tikzpicture}[scale=0.8]
      \draw[very thick] (0,0) ellipse(5cm and 3cm);
      \foreach \x/\xlabel in {-2.5/M_i,2.5/M_j}
        \draw[very thick] (\x,-0.5) node[below,yshift=-1cm]{$\xlabel$}
          ellipse(1cm and 0.75cm);
      \draw[very thick] (2.75,-0.35) arc(10:170:2.75cm);
      \draw[very thick] (2.25,-0.65) arc(10:170:2.25cm);
      \draw[very thick] (0.4,1.55) arc(-40:-140:0.5cm);
      \draw[very thick] (0.295,1.45) arc(30:150:0.325cm);
      \node at (4,2.75){$B$};
    \end{tikzpicture}
  } \vspace*{1cm} \subfigure[Glued]{
    \begin{tikzpicture}[scale=0.8]
      \draw[very thick] (0,0) ellipse(5cm and 3cm);
      \node at (4,2.75){$B$};
      \draw[very thick] (-2.5,1.5) arc(90:270:1.5cm);
      \draw[very thick] (2.5,-1.5) arc(-90:90:1.5cm);
      \draw[very thick] (2.5,1.5) .. controls (1.25,1.5) and (1.25,0.75)
        .. (0,0.75) .. controls (-1.25,0.75) and (-1.25,1.5)
        .. (-2.5,1.5);
      \draw[very thick] (2.5,-1.5) .. controls (1.25,-1.5) and
        (1.25,-0.75) .. (0,-0.75) .. controls (-1.25,-0.75) and
        (-1.25,-1.5) .. (-2.5,-1.5);
      \draw[very thick] (0,0) ellipse(0.25cm and 0.75cm);
      \node at (-2.5,0){$M_i$};
      \node at (2.5,0){$M_j$};
    \end{tikzpicture}
  }
  \caption{}
  \label{fig:manifold-bonds}
\end{figure}

\setcounter{subfigure}{0}

\begin{figure}[H]
  \centering
  \subfigure[Cobordant]{        
    \begin{tikzpicture}
      \draw[very thick] (0,0) node{$S$} ellipse(4cm and 3.5cm);
      \foreach \a in {-2,2}{
        \draw[very thick] (\a,0) ellipse(0.25cm and 0.5cm);
        \draw[very thick] (0,\a) ellipse(0.5cm and 0.25cm);
      }
      \draw[very thick] (2,0.5) .. controls (0.5,0.5) .. (0.5,2);
      \draw[very thick] (2,-0.5) .. controls (0.5,-0.5) .. (0.5,-2);
      \draw[very thick] (-2,0.5) .. controls (-0.5,0.5) .. (-0.5,2);
      \draw[very thick] (-2,-0.5) .. controls (-0.5,-0.5) .. (-0.5,-2);
      \node[left] at (-2.25,0){$S_1$};
      \node[above] at (0,2.25){$S_2$};
      \node[below] at (0,-2.25){$S_3$};
      \node[right] at (2.25,0){$S_4$};
      \node at (3.5,3){$B$};
      \node[below] at (0,-3.75){$P_i \to S_i, \qquad \coprod S_i
        = \partial S \subset B \quad \text{or} \quad S = B$};
    \end{tikzpicture}
  } \vspace*{0.25cm} \subfigure[Weakly cobordant]{
    \begin{tikzpicture}
      \draw[very thick] (0,0) node{$S$} ellipse(4cm and 3.5cm);
      \draw[very thick] (2,0.5) .. controls (0.5,0.5) .. (0.5,2) --
        (-0.5,2) .. controls (-0.5,0.5) .. (-2,0.5) -- (-2,-0.5)
        .. controls (-0.5,-0.5) .. (-0.5,-2) -- (0.5,-2) .. controls
        (0.5,-0.5) .. (2,-0.5) -- (2,0.5);
      \node[left] at (-2.25,0){$S_1$};
      \node[above] at (0,2.25){$S_2$};
      \node[below] at (0,-2.25){$S_3$};
      \node[right] at (2.25,0){$S_4$};
      \node at (3.5,3){$B$};
      \node[below] at (0,-3.75){$\coprod S_i \subset \partial S
        \subset B$};
    \end{tikzpicture}
  }
  \caption{}
  \label{fig:cobordant-bonds}
\end{figure}

\newpage

\begin{example} $B = \R^3$
  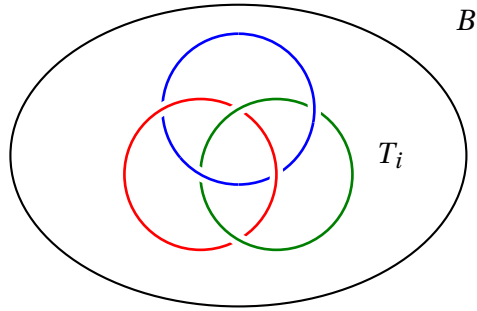
\begin{figure}[H]
    \centering
    \begin{tikzpicture}
      \draw[thick] (0,0) ellipse (3cm and 2cm);
      \draw (2,0) node{$T_i$};
      \draw (3,1.8) node{$B$};

      \begin{scope}[scale=0.5,every
        path/.style={knot},xshift=-1cm,yshift=-0.5cm]
        \pgfmathsetmacro{\csixty}{cos(60)}
        \pgfmathsetmacro{\ssixty}{sin(60)}
        \pgfmathsetmacro{\brscale}{2}
        \draw[double=ring1] (0,0) circle (\brscale);
        \draw[double=ring2] (\brscale,0) circle (\brscale);
        \begin{pgfonlayer}{back}
          \draw[double=ring3] (\brscale *\csixty,\brscale *\ssixty)
            ++(0,\brscale) arc(90:220:\brscale);
          \draw[double=ring3] (\brscale *\csixty,\brscale *\ssixty)
            ++(0,-\brscale) arc(-90:-10:\brscale);
        \end{pgfonlayer}
        \draw[double=ring3] (\brscale *\csixty,\brscale *\ssixty)
          ++(0,\brscale) arc(90:-10:\brscale);
        \draw[double=ring3] (\brscale *\csixty,\brscale *\ssixty)
          ++(0,-\brscale) arc(-90:-140:\brscale);
      \end{scope}
    \end{tikzpicture}
    \caption{Borromean interactions}
    \label{fig:BorroInt}
  \end{figure}

  \begin{figure}[H]
    \centering
    \begin{tikzpicture}
      \draw[thick] (0,0) ellipse (3cm and 2cm);
      \draw (2,0) node{$T_i$};
      \draw (3,1.8) node{$B$};

      \begin{scope}[scale=0.5,every
        path/.style={knot},xshift=-8cm,yshift=-0.75cm]
        \pgfmathsetmacro{\ssixty}{sin(60)}
        \pgfmathsetmacro{\brscale}{2}
        \pgfmathsetmacro{\yshift}{\brscale * \ssixty/2}
        \pgfmathsetmacro{\brscale}{1.8}
        \begin{scope}[xshift=4.5*\brscale cm,yshift=\yshift cm]
          \foreach \brk in {1,2,3} {
            \begin{scope}[rotate=\brk * 120 - 120]
              \colorlet{chain}{ring\brk}
              \brunnianlink{\brscale}{120}
            \end{scope}
          }
        \end{scope}
      \end{scope}
    \end{tikzpicture}
    \caption{Brunnain interactions}
    \label{fig:BrunnInt}
  \end{figure}
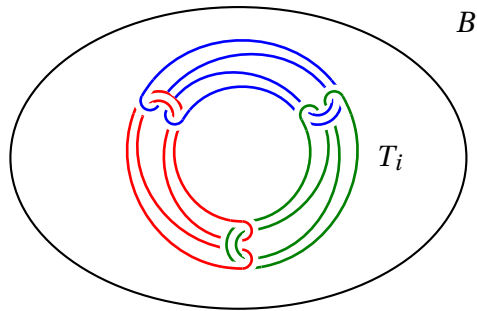

  In this paper we have discussed a whole hierarchy of extensions.
\end{example}

For higher dimensional manifolds there is a variety of ways to do
this (spaces of embeddings).

One may manipulate or tune externally a system in such a way that it
is represented by a desired (linked,$\ldots$) embedding of the $M_i$'s
in $B$.

Hence it justifies calling $B$ a bond, see \citeasnoun{Baas}.\\

Then one may --- as already mentioned --- iterate this process to
higher order bonds which we have defined as hyperstructures.  This
gives higher order, hyperstructured interaction patterns of the
particles.\\

\emph{This means that for many-body systems there is a whole new
universe of new represented states}.

\emph{The pertinent question is then: Which of these new types of
  states are realizable in real world systems (physical, chemical,
  biological, social,$\ldots$)?}\\

This discussion also shows that bonds of subspaces (like manifolds)
and their associated hyperstructures may be a good laboratory for
suggesting new states, designing and studying general many-body
systems and their interactions.  But the geometric interactions in the
geometric universe should then be interpreted back into interactions
in the real universe where the particles live.  In cold gases
Borromean or Brunnian states of first order correspond to Efimov
states as explained in the main text where we have studied and
suggested connections between physical states and higher order links
in the geometric universe of links.  For more general types of
many-body systems, see \citeasnoun{structure}.

\section*{Appendix}

In this appendix we show a wide variety of first, second and third
order links.\\

\appendix

\section{A Hopf family}
\begin{figure}[H]
  \centering
  \includegraphics[width=0.5\linewidth]{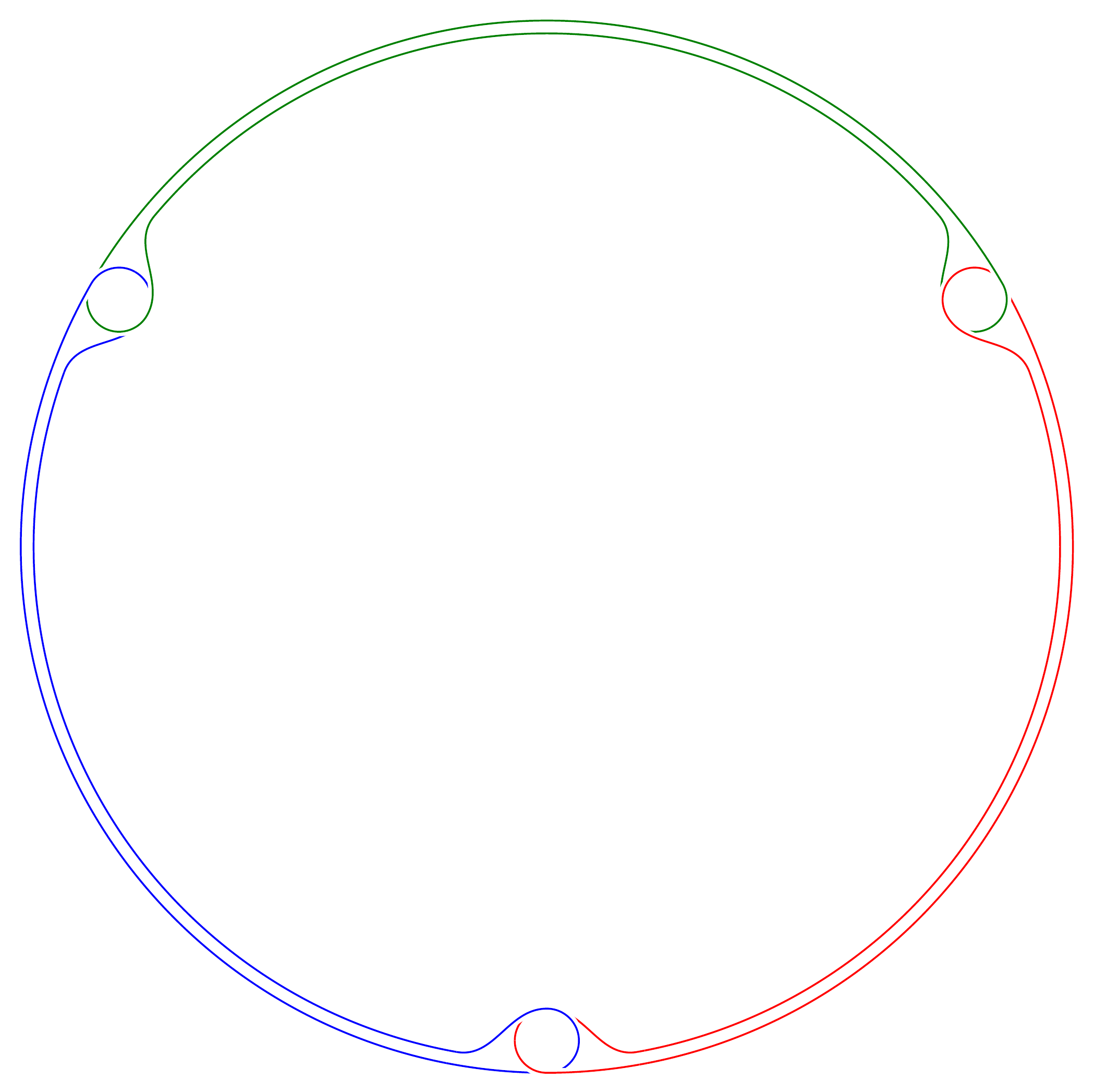}
  \caption{Type $1H(3)$}
  \label{fig:1h3}
\end{figure}

\begin{figure}[H]
  \centering
  \includegraphics[width=0.5\linewidth]{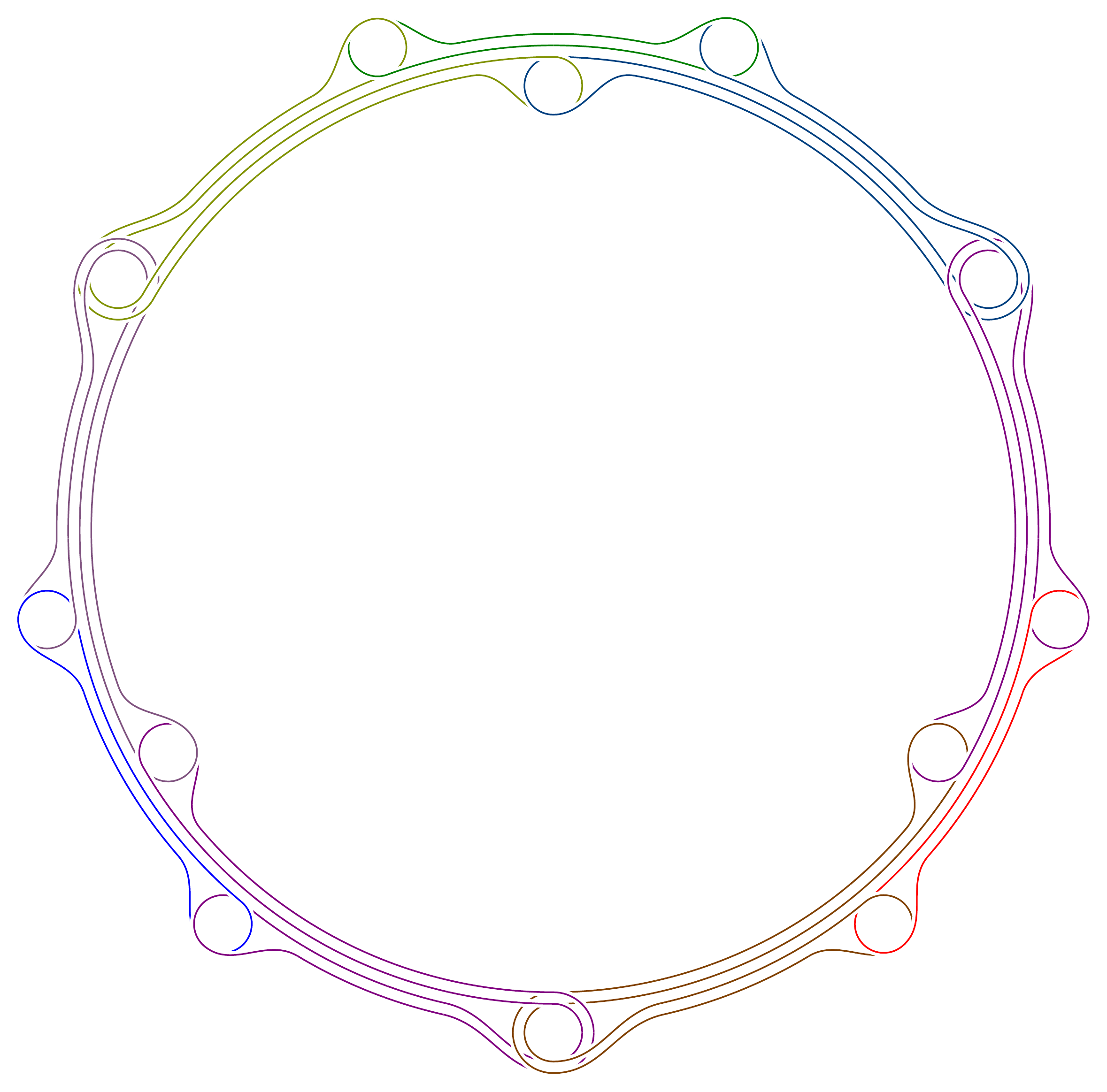}
  \caption{Type $2H(3,3)$}
  \label{fig:2h33}
\end{figure}

\begin{figure}[H]
  \centering
  \includegraphics[width=0.5\linewidth]{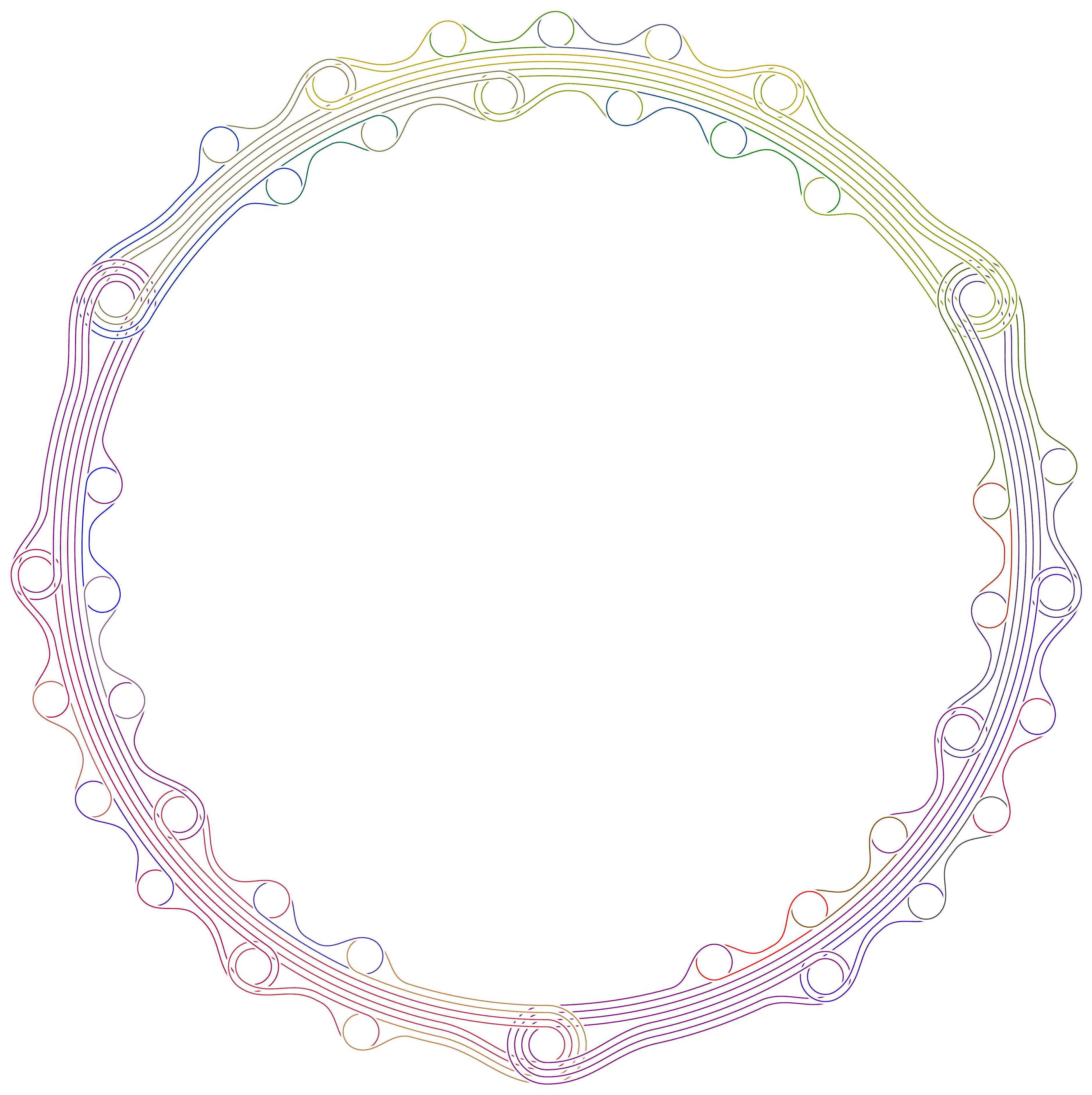}
  \caption{Type $3H(3,3,3)$}
  \label{fig:3h333}
\end{figure}

\begin{figure}[H]
  \centering
  \includegraphics[width=\ScaleIfNeeded]{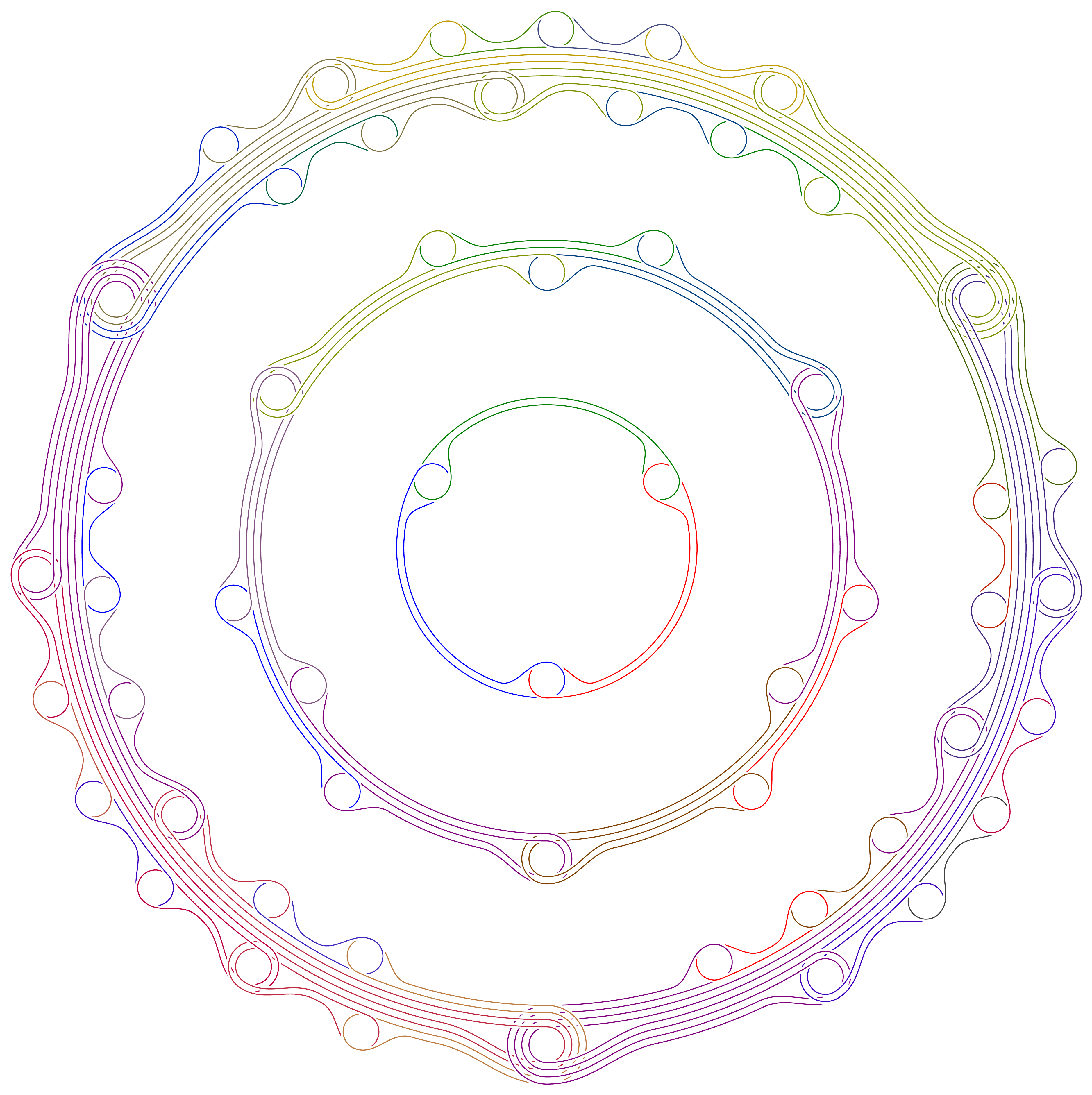}
  \caption{Inner ring: $1H(3)$ \quad Middle ring: $2H(3,3)$ \quad
    Outer ring: 3H(3,3,3)}
  \label{fig:}
\end{figure}

\section{A Brunnian family}

\begin{figure}[H]
  \centering
  \includegraphics[width=0.475\linewidth]{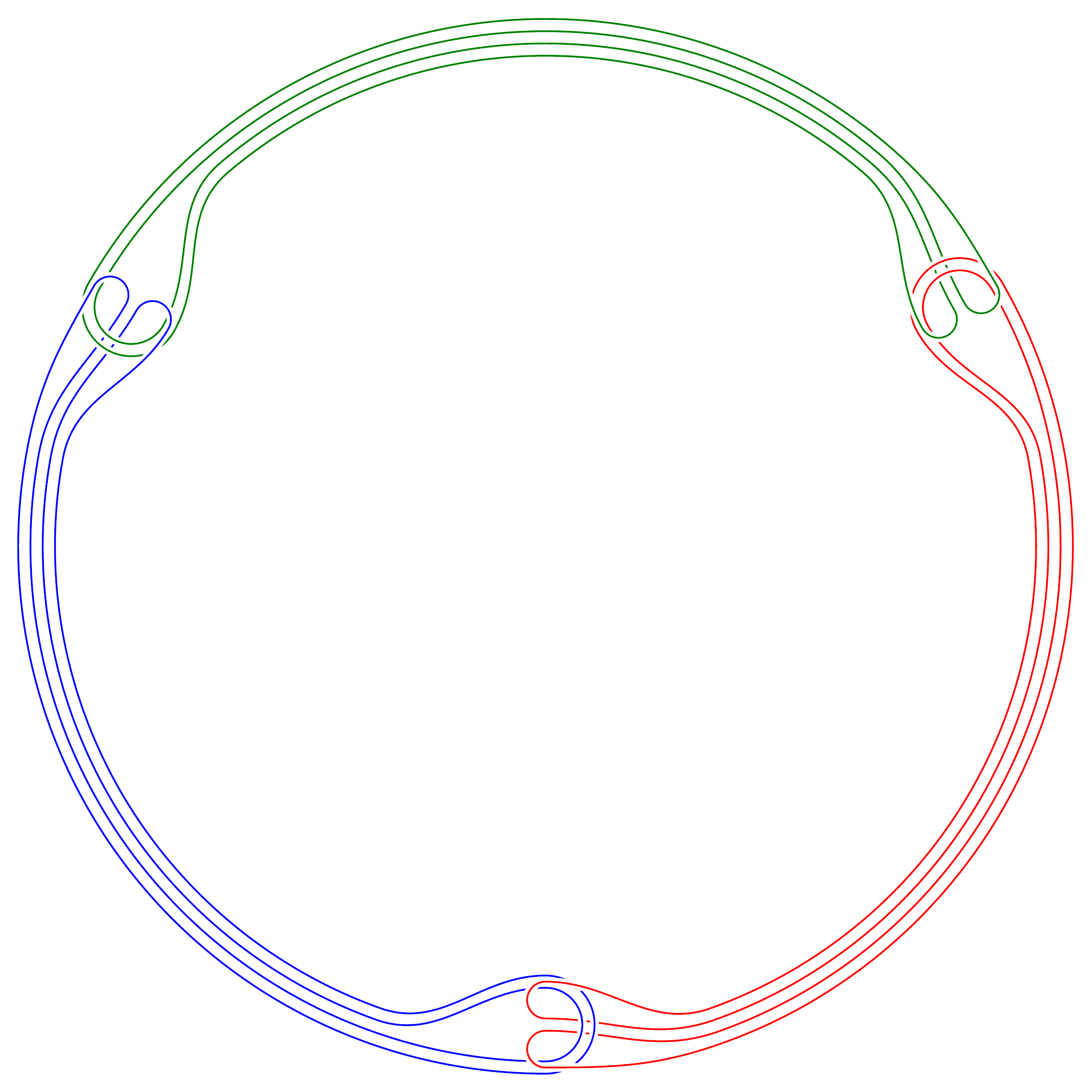}
  \caption{Type $1B(3)$}
  \label{fig:1b3}
\end{figure}

\begin{figure}[H]
  \centering
  \includegraphics[width=0.475\linewidth]{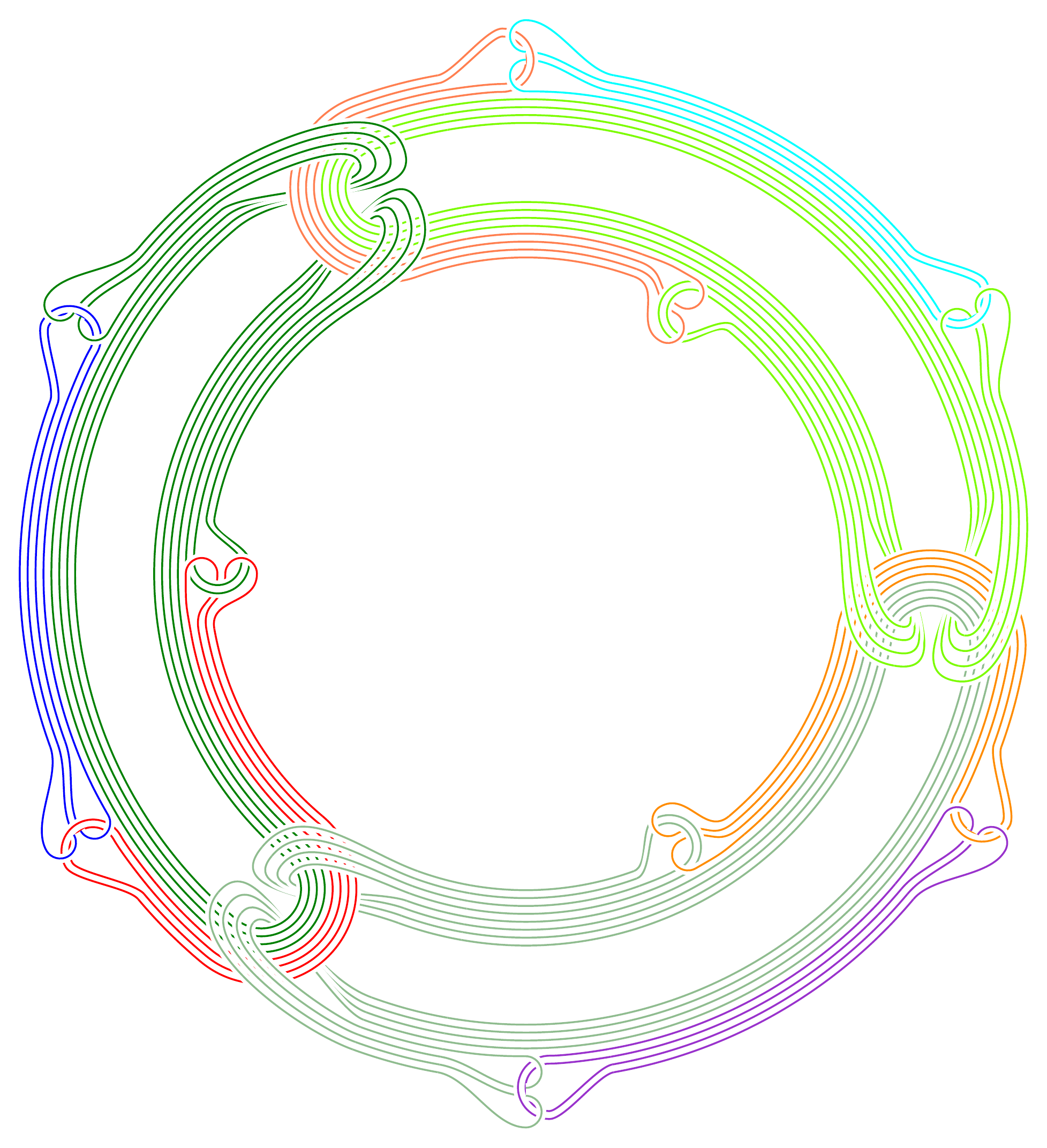}
  \caption{Type $2B(3,3)$}
  \label{fig:2b33}
\end{figure}

\begin{figure}[H]
  \centering
  \includegraphics[width=0.6\linewidth]{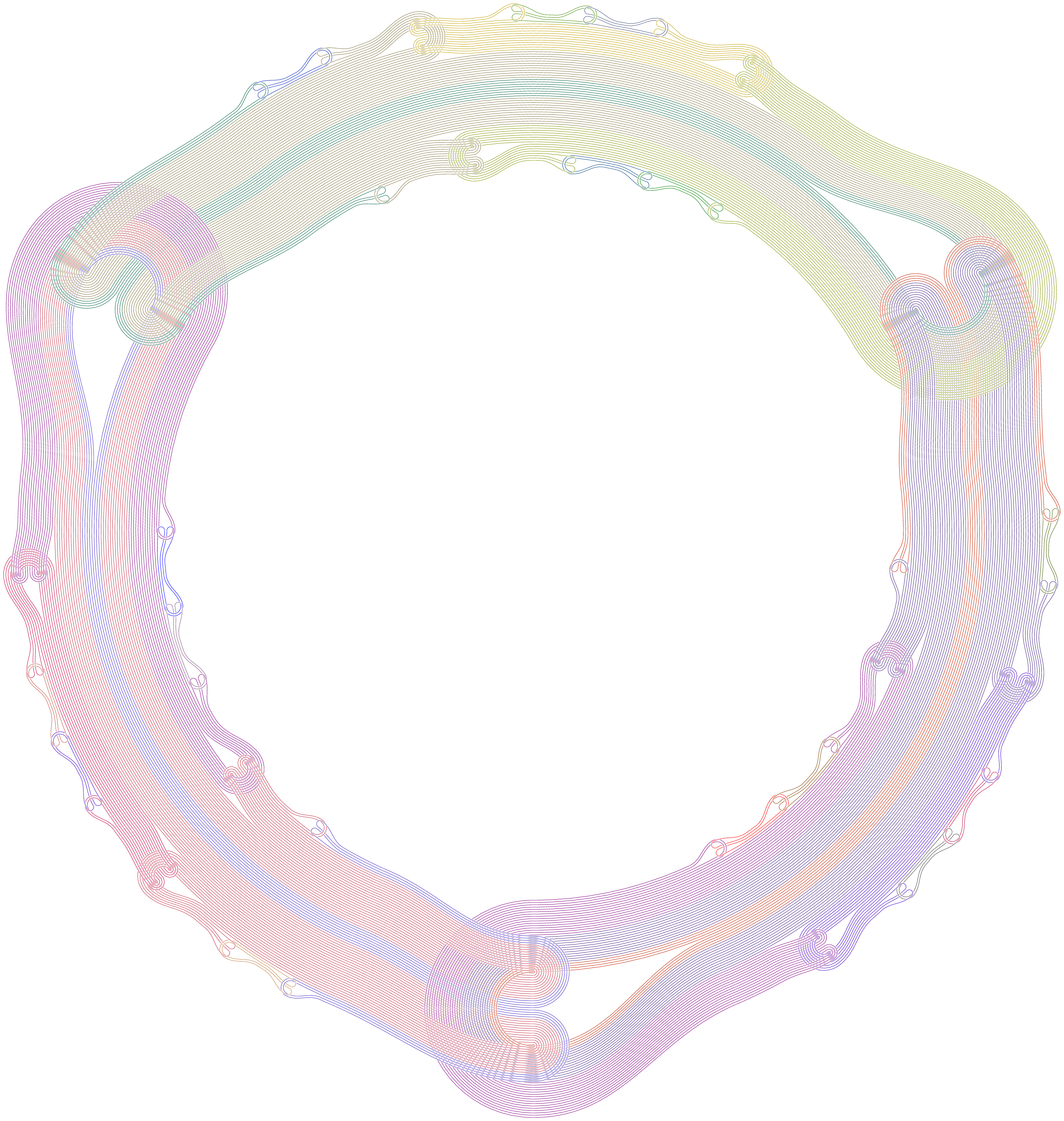}
  \caption{Type $3B(3,3,3)$}
  \label{fig:3b333}
\end{figure}

\begin{figure}[H]
  \centering
  \includegraphics[width=0.6\linewidth]{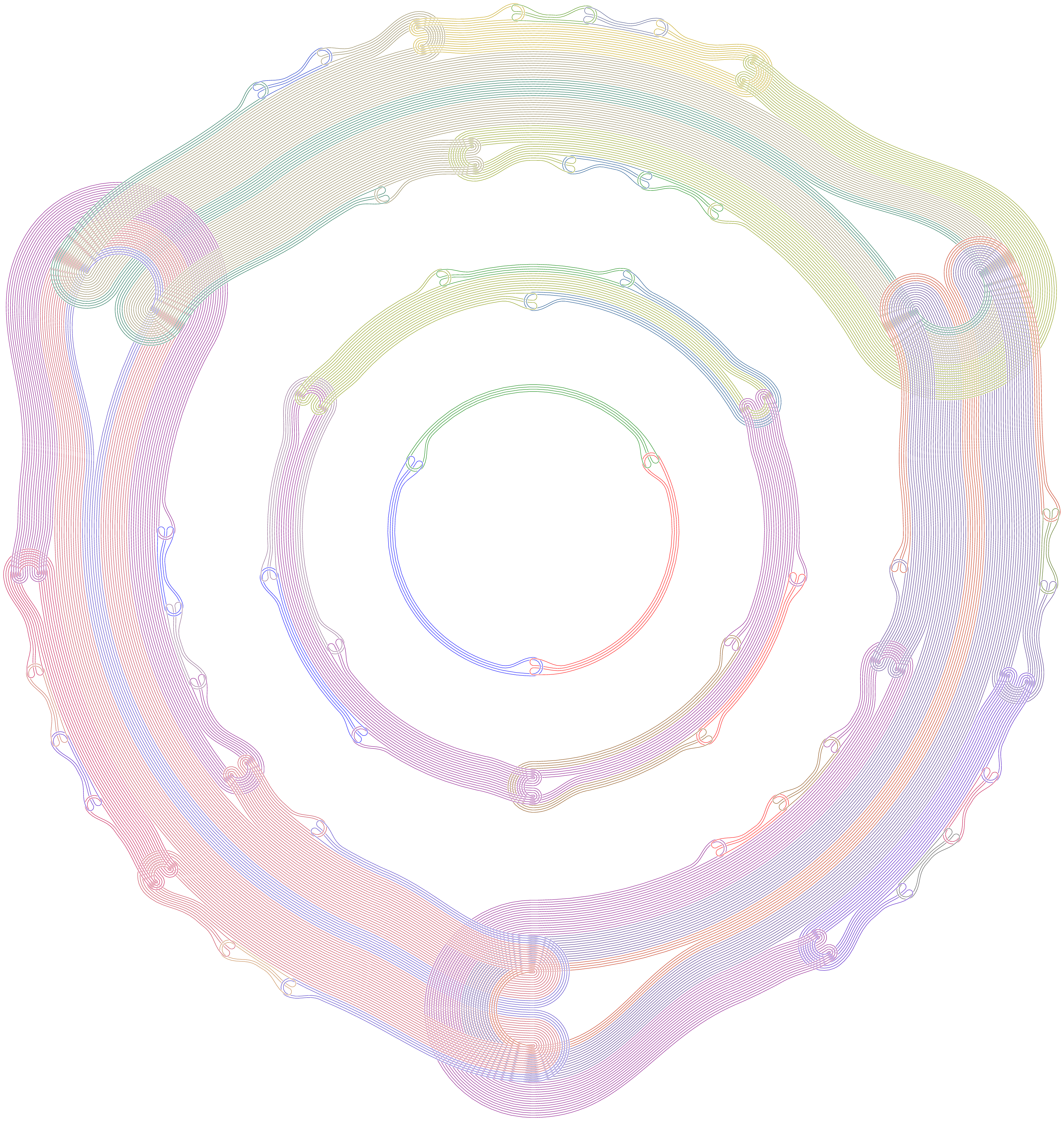}
  \caption{Inner ring: $1B(3)$ \quad Middle ring: $2B(3,3)$ \quad
    Outer ring: $3B(3,3,3)$}
  \label{fig:}
\end{figure}

\section{A two-ring family}

\begin{figure}[H]
  \centering
  \includegraphics[width=0.5\linewidth]{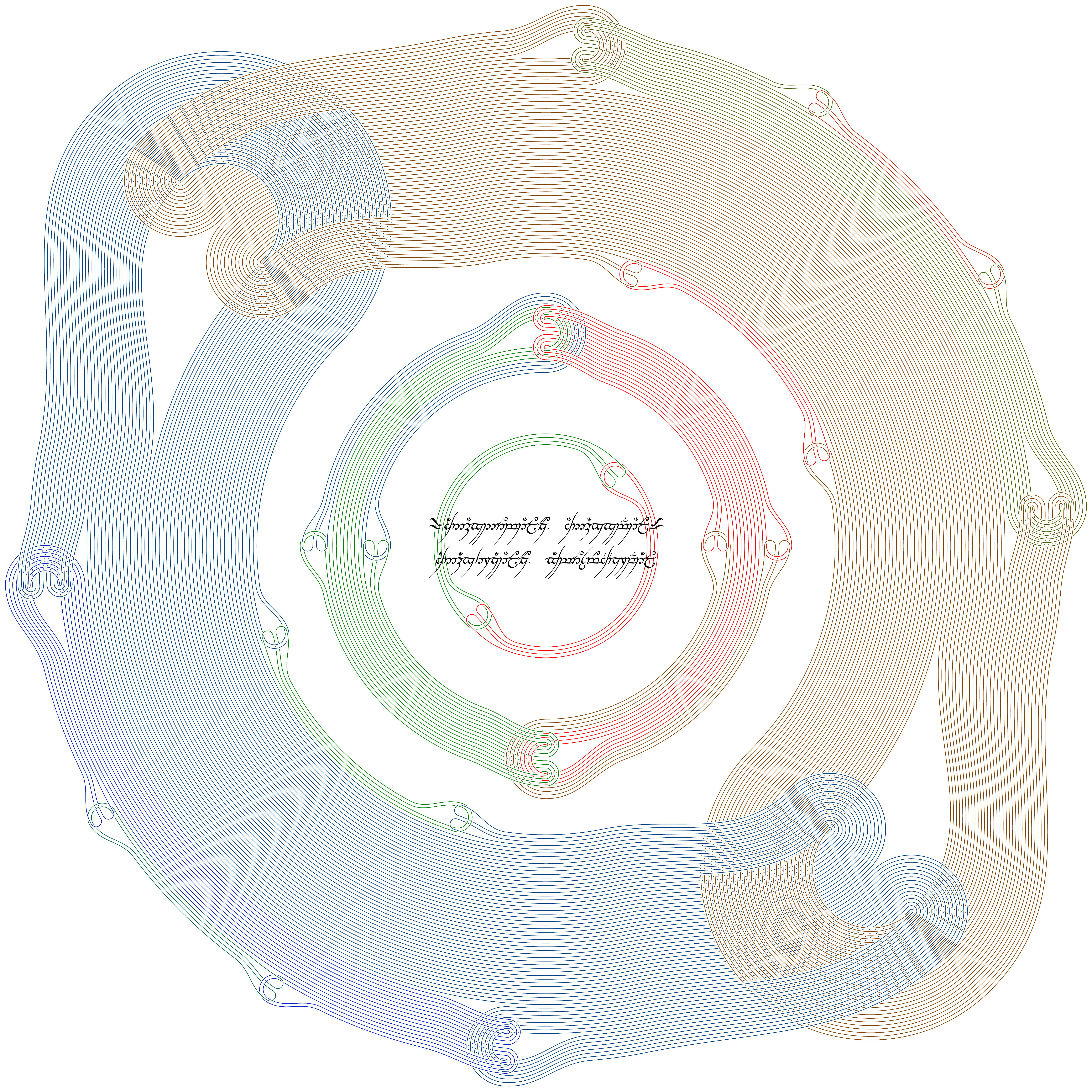}
  \caption{Inner ring: $1B(2)$ \quad Middle ring: $2B(2,2)$ \quad
    Outer ring: $3B(2,2,2)$ The inscription is from J.R.R.\ Tolkien,
    \emph{The Lord of the Rings}: ``One Ring to rule them all, One
    Ring to find them, One Ring to bring them all and in the darkness
    bind them''}
\end{figure}

\section{Various second order Brunnian examples}

\begin{figure}[H]
  \centering
  \includegraphics[width=0.5\linewidth]{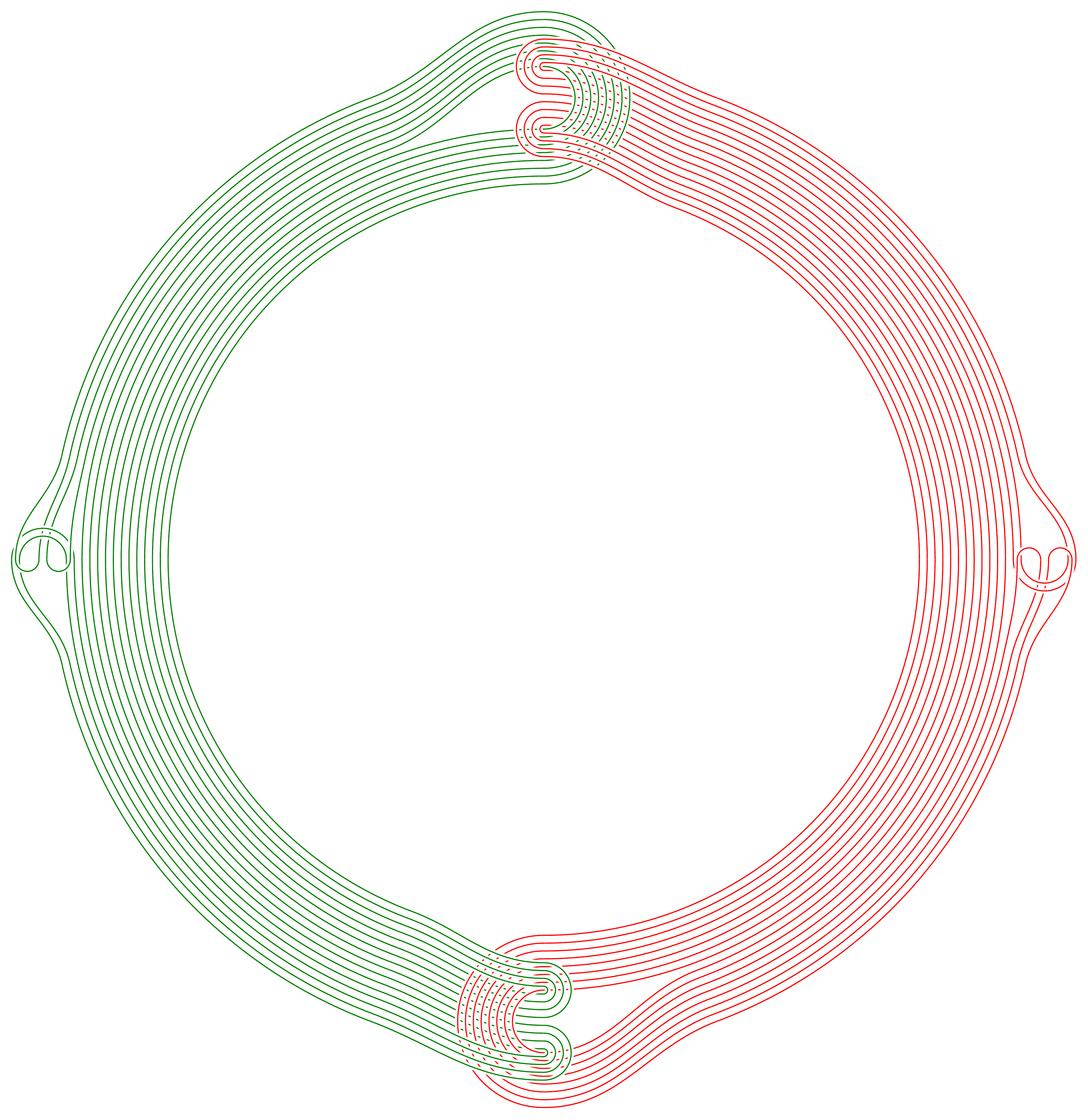}
  \caption{Type $2B(1,2)$}
  \label{fig:}
\end{figure}

\begin{figure}[H]
  \centering
  \includegraphics[width=0.6\linewidth]{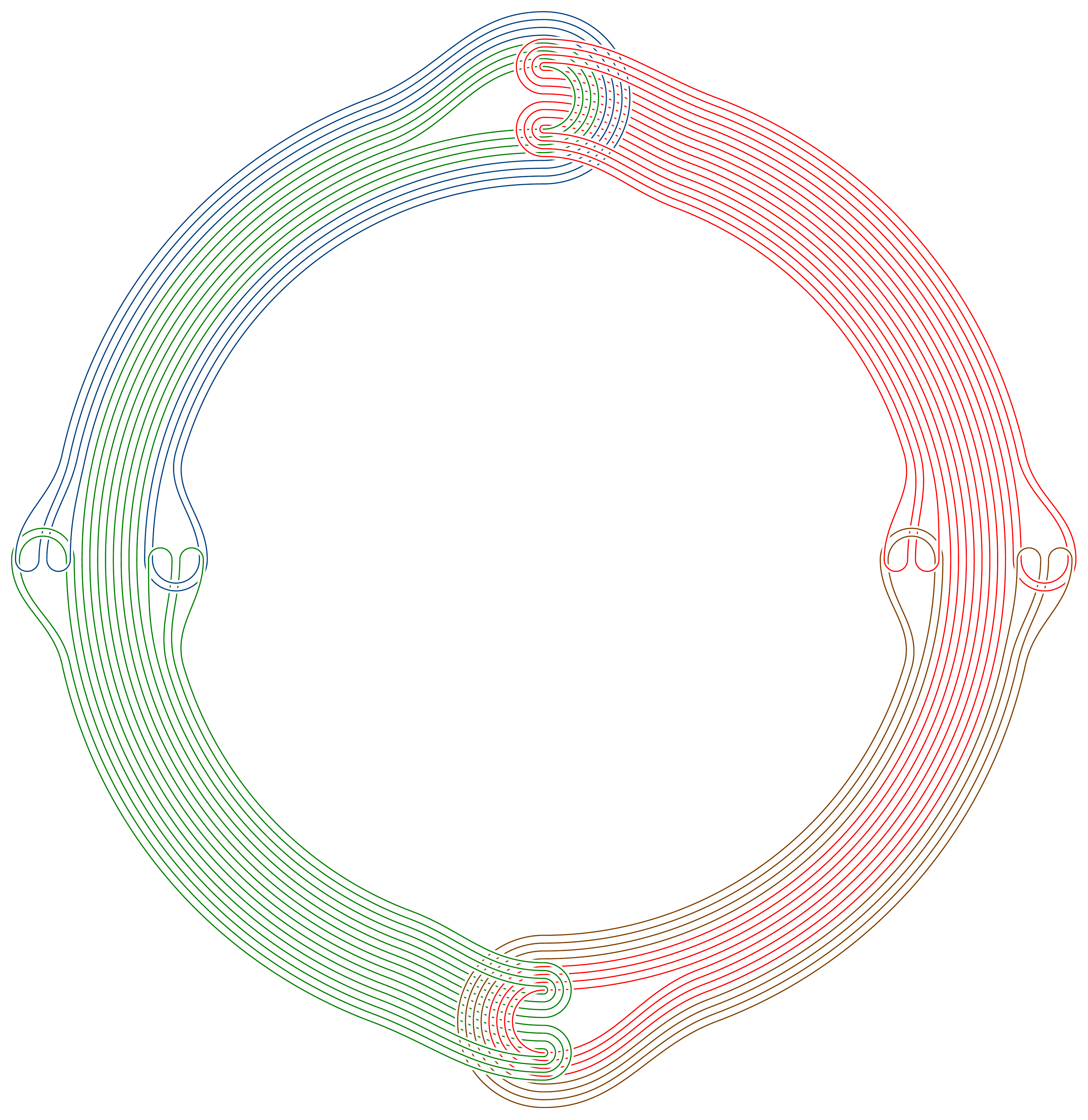}
  \caption{Type $2B(2,2)$}
  \label{fig:}
\end{figure}

\begin{figure}[H]
  \centering
  \includegraphics[width=0.6\linewidth]{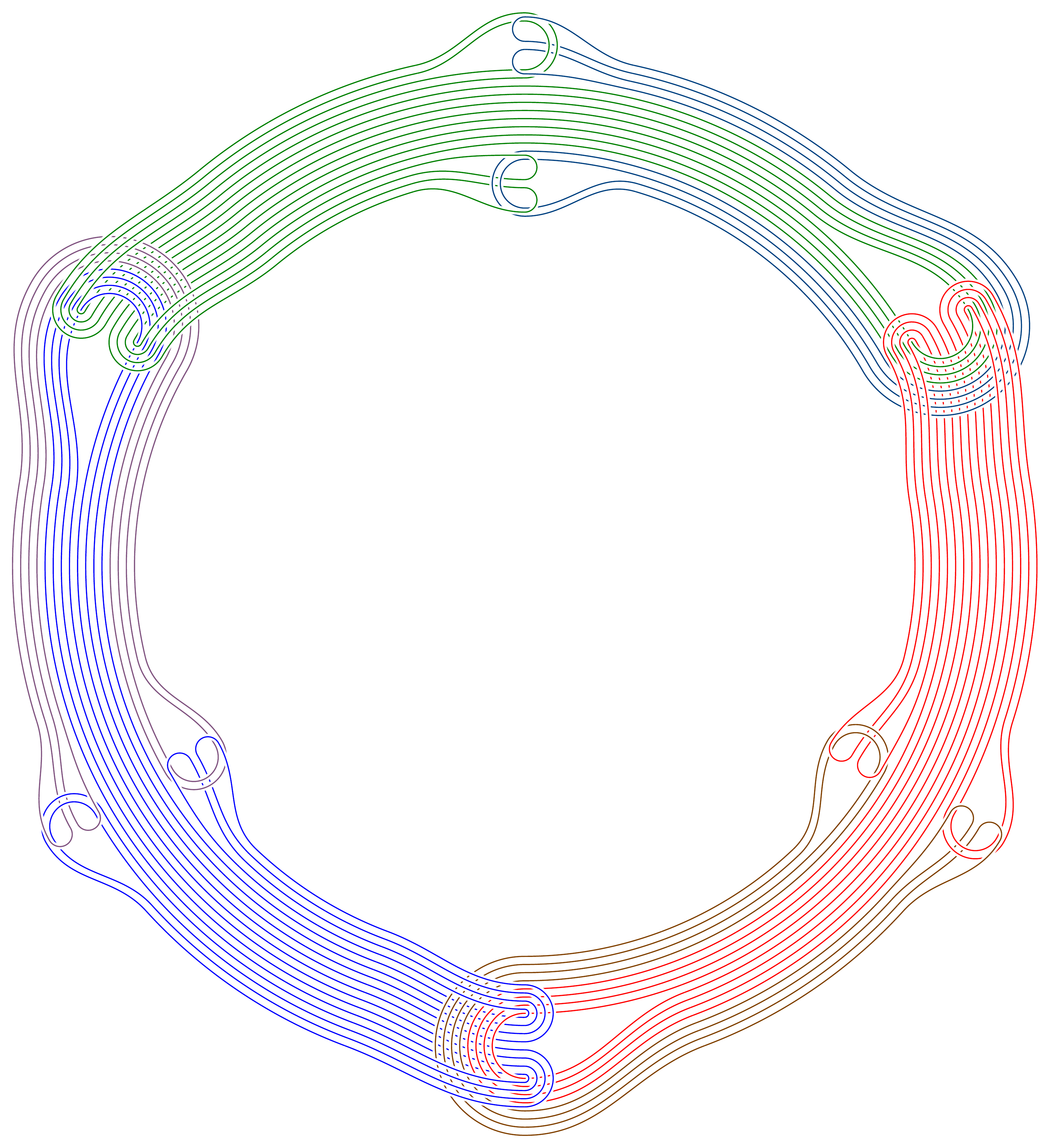}
  \caption{Type $2B(2,3)$}
  \label{fig:}
\end{figure}

\begin{figure}[H]
  \centering
  \includegraphics[width=0.6\linewidth]{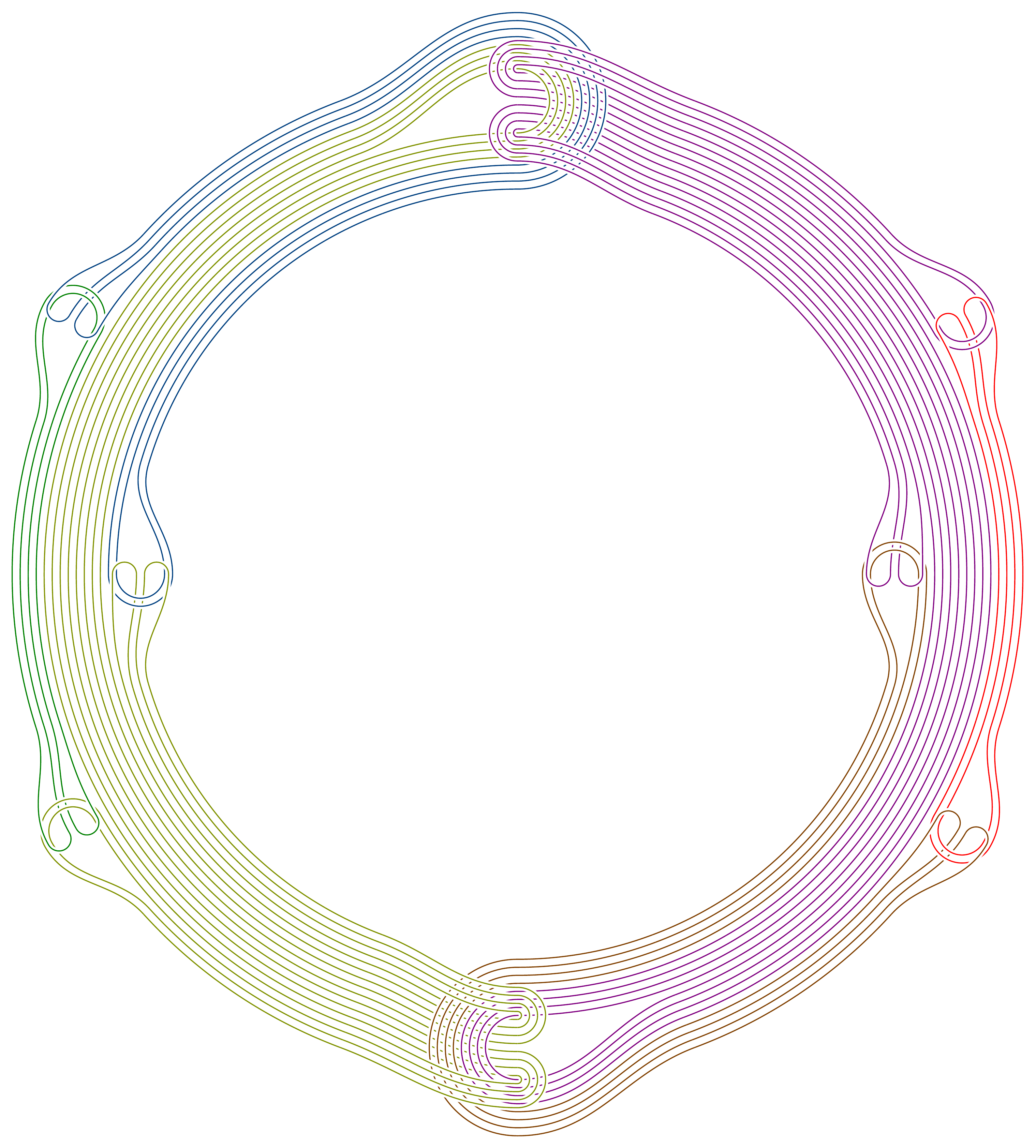}
  \caption{Type $2B(3,2)$}
  \label{fig:}
\end{figure}

\begin{figure}[H]
  \centering
  \includegraphics[width=0.6\linewidth]{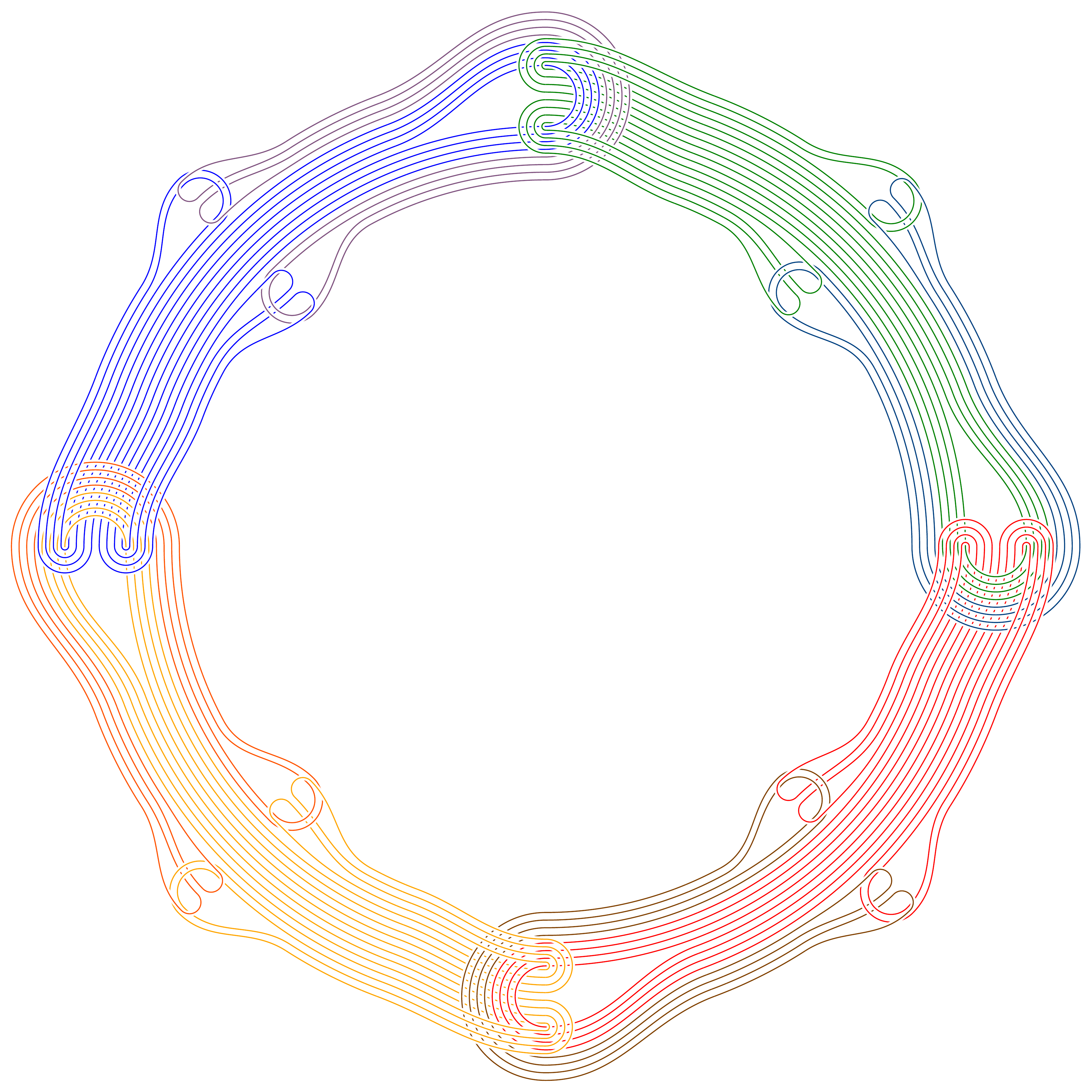}
  \caption{Type $2B(2,4)$}
  \label{fig:}
\end{figure}

\begin{figure}[H]
  \centering
  \includegraphics[width=0.6\linewidth]{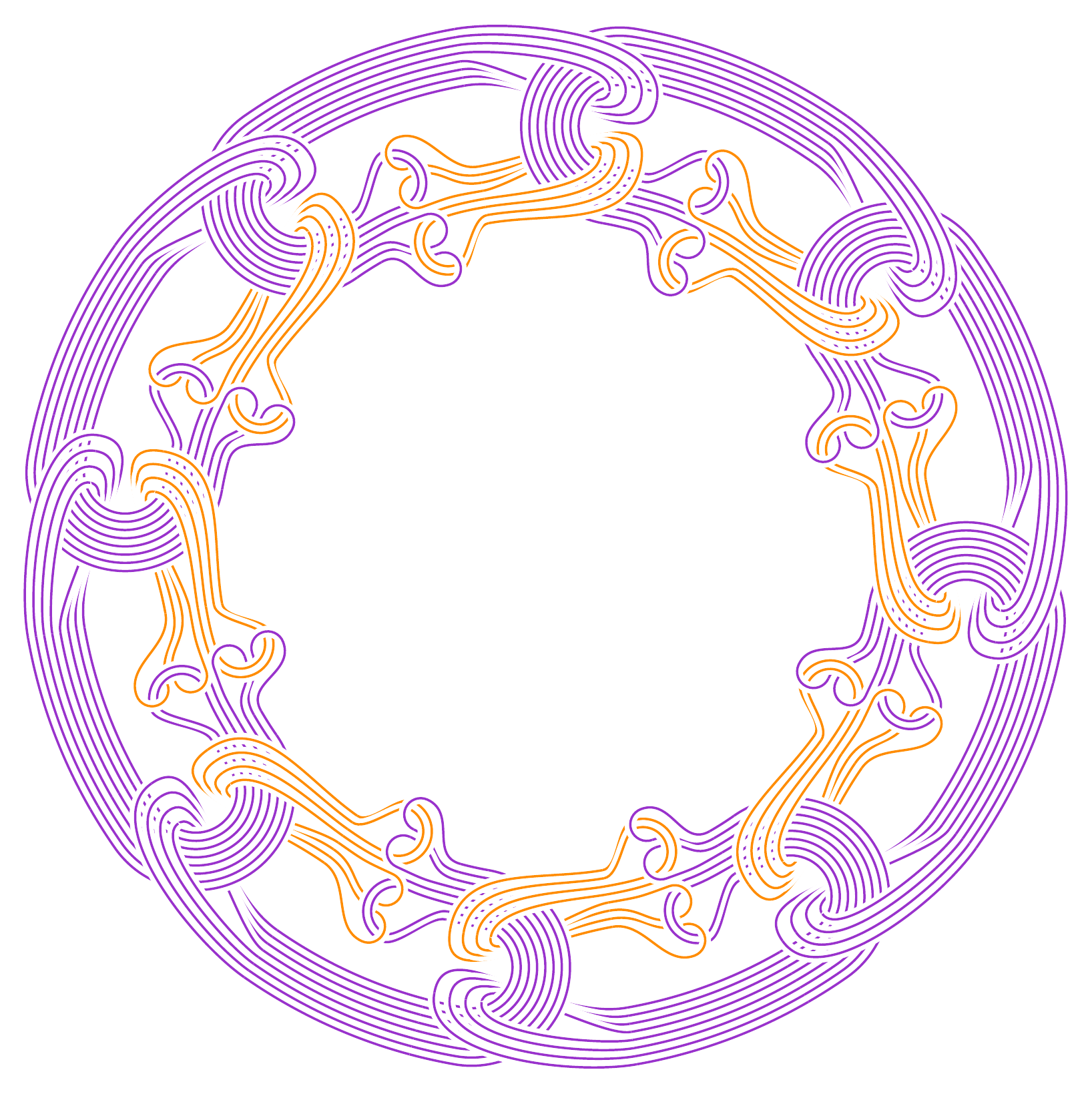}
  \caption{Type $2B(2,8)$}
  \label{fig:}
\end{figure}

\begin{figure}[H]
  \centering
  \includegraphics[width=0.6\linewidth]{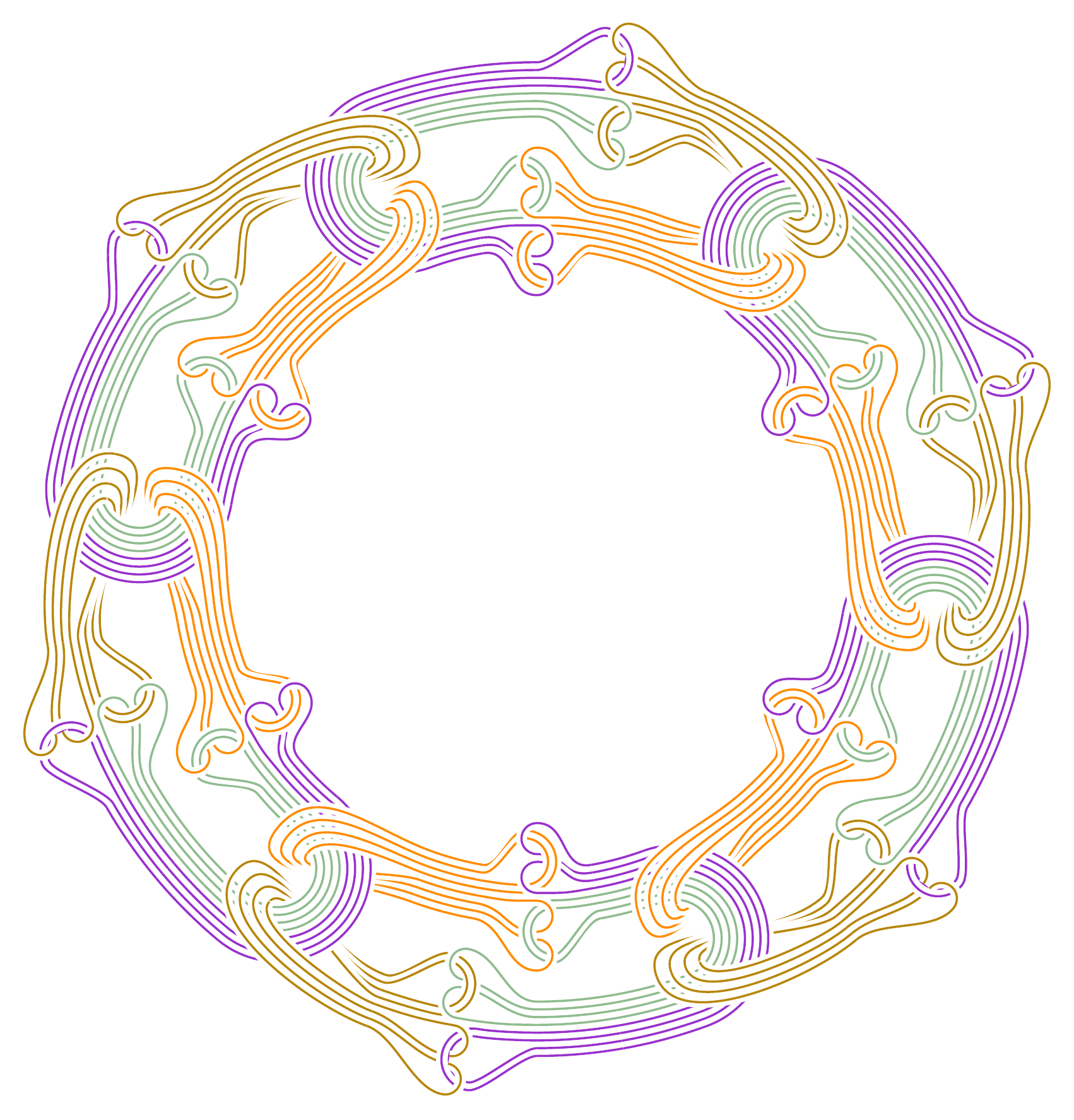}
  \caption{Type $2B(4,6)$}
  \label{fig:}
\end{figure}

\section{Borromean and Brunnian rings}
In this section we show that the Borromean and Brunnian rings in
Figure \ref{fig:1} are not equivalent (isotopic) through the
following deformations:

\begin{figure}[H]
  \centering
  \begin{tikzpicture}[scale=0.5]
    \pgfmathsetmacro{\csixty}{cos(60)}
    \pgfmathsetmacro{\ssixty}{sin(60)}
    \pgfmathsetmacro{\brscale}{1}
    \begin{scope}[xshift=.5 * \brscale cm]
      \draw[knot,double=ring1] (0,0) circle (\brscale);
      \draw[knot,double=ring2] (\brscale,0) circle (\brscale);
      \begin{pgfonlayer}{back}
        \draw[knot,double=ring3] (\brscale *\csixty,\brscale *\ssixty)
          ++(0,\brscale) arc(90:220:\brscale);
        \draw[knot,double=ring3] (\brscale *\csixty,\brscale *\ssixty)
          ++(0,-\brscale) arc(-90:-10:\brscale);
      \end{pgfonlayer}
      \draw[knot,double=ring3] (\brscale *\csixty,\brscale *\ssixty)
        ++(0,\brscale) arc(90:-10:\brscale);
      \draw[knot,double=ring3] (\brscale *\csixty,\brscale *\ssixty)
        ++(0,-\brscale) arc(-90:-140:\brscale); 
    \end{scope}

    \pgfmathsetmacro{\brxshift}{6 + .75 * \brscale}
    \begin{scope}[xshift= \brxshift cm]
      \draw[knot,double=ring1] (0,0) circle (\brscale);
      \draw[knot,double=ring2] (1.6 * \brscale,0) circle (\brscale);
      \begin{pgfonlayer}{back}
        \draw[knot,double=ring3] (\brscale *\csixty,\brscale *\ssixty)
          ++(0,\brscale) arc(90:270:\brscale);
        \draw[knot,double=ring3] (\brscale *\csixty,\brscale *\ssixty)
          ++(\brscale,0) arc(0:-70:\brscale);
      \end{pgfonlayer}
      \draw[knot,double=ring3] (\brscale *\csixty,\brscale *\ssixty)
        ++(0,\brscale) arc(90:0:\brscale);
      \draw[knot,double=ring3] (\brscale *\csixty,\brscale *\ssixty)
        ++(0,-\brscale) arc(-90:-70:\brscale);
    \end{scope}

    \begin{scope}[yshift=-4cm]
      \draw[knot,double=ring1] (0,0) circle (\brscale);
      \draw[knot,double=ring2] (2*\brscale,0) circle (\brscale);
      \draw[knot,double=ring3] (\brscale *\csixty,\brscale *\ssixty)
        ++(0,\brscale) arc(90:-10:\brscale) node[coordinate] (a) {};
      \draw[knot,double=ring3] (\brscale *\csixty,\brscale *\ssixty)
        ++(0,-\brscale) arc(-90:-40:\brscale) node[coordinate] (b) {};
      \node[coordinate] (e) at (\brscale,.65*\brscale) {};
      \begin{pgfonlayer}{back}
        \draw[knot,double=ring3] (\brscale *\csixty,\brscale *\ssixty)
          ++(0,\brscale) arc(90:270:\brscale);
        \draw[knot,double=ring3] (\brscale *\csixty,\brscale *\ssixty)
          ++(0,.5*\brscale) node[coordinate] (c) {}
          arc(90:270:.5*\brscale) node[coordinate] (d) {};
        \draw[knot,double=ring3] (a) to[out=260,in=0] (c);
        \draw[knot,double=ring3] (b) to[out=50,in=0] (e);
      \end{pgfonlayer}
      \draw[knot,double=ring3] (e) to[out=180,in=0] (d);
    \end{scope}

    \begin{scope}[yshift=-4cm, xshift=7.5cm]
      \draw[knot,double=ring1] (-1.3*\brscale0,0) circle (\brscale);
      \draw[knot,double=ring2] (1.5*\brscale,0) circle (\brscale);
      \colorlet{chain}{ring3}
      \flatbrunnianlink{\brscale}
    \end{scope}

    \draw[ultra thick,black,double=none,->] (3.25*\brscale,0) --
      ++(2*\brscale,0);
    \draw[ultra thick,black,double=none,->] (3.5*\brscale,-4) --
      ++(1.25*\brscale,0);
    \draw[ultra thick,black,double=none,->] (5.25*\brscale,-1) --
      (3.5*\brscale,-3);
  \end{tikzpicture}
  \caption{}
  \label{fig:37}
\end{figure}

\begin{figure}[H]
  \centering
  \begin{tikzpicture}[scale=0.6]
    \pgfmathsetmacro{\brscale}{1}
    \begin{scope}[xshift=6.5 cm]
      \draw[knot,double=ring1] (-6.3*\brscale0,0) circle (\brscale);
      \draw[knot,double=ring2] (-1.5*\brscale,0) circle (\brscale);
      \foreach \brk in {5,...,3} {
        \begin{scope}[xshift=-\brk * \brscale cm]
          \colorlet{chain}{ring\brk}
          \flatbrunnianlink{\brscale}
        \end{scope}
      }
    \end{scope}

    \begin{scope}[yshift=-4cm,xshift=-1.375*\brscale cm]
      \setbrstep{.125}
      \draw[knot,double=ring1] (\brscale,0) arc(0:-180:\brscale);
      \draw[knot,double=ring2] (7.75*\brscale,0) circle (\brscale);
      \foreach \brk in {3,...,6} {
        \pgfmathparse{\brk == 3 ? "white" : "ring"}
        \edef\minner{\pgfmathresult}
        \pgfmathparse{\brk == 3 ? "" : int(8 - \brk + 1)}
        \edef\minner{\minner\pgfmathresult}
        \pgfmathparse{\brk == 6 ? "white" : "ring"}
        \edef\mouter{\pgfmathresult}
        \pgfmathparse{\brk == 6 ? "" : int(8 - \brk)}
        \edef\mouter{\mouter\pgfmathresult}
        \pgfmathsetmacro{\brl}{(2*\brk - 5)*\brscale}
        \begin{scope}[xshift=\brl cm]
          \colorlet{inner}{\minner}
          \colorlet{outer}{\mouter}
          \flatjunction{1}{10}{-1}{}{}
        \end{scope}
      }
      \draw[knot,double=ring1] (\brscale,0) arc(0:180:\brscale);
    \end{scope}

    \begin{scope}[yshift=-10cm,xshift=2.5*\brscale cm]
      \setbrstep{.125}
      \draw[knot,double=ring2] (.7,3) circle (\brscale);
      \draw[knot,double=ring1] (3,0) arc(270:90:\brscale);
      \begin{scope}
        \colorlet{ring1}{ring5}
        \colorlet{ring2}{white}
        \brunnian{3}{4}
      \end{scope}
      \draw[knot,double=ring1] (3,0) arc(-90:90:\brscale);
    \end{scope}

    \begin{scope}[yshift=-18cm,xshift=2.5*\brscale cm]
      \setbrstep{.125}
      \colorlet{ring1}{ring5}
      \colorlet{ring2}{ring3}
      \colorlet{ring3}{ring4}
      \brunnian{3}{3}
    \end{scope}
  \end{tikzpicture}
  \caption{}
  \label{fig:38}
\end{figure}

Comparing the end results of Figures \ref{fig:37} and \ref{fig:38}
shows that they are not isotopic.  (A calculation of the Jones
polynomials proves this.)\\

The following two links are Borromean of length $4$, but not Brunnian.

\begin{figure}[H]
  \centering
  \includegraphics[width=\ScaleIfNeeded]{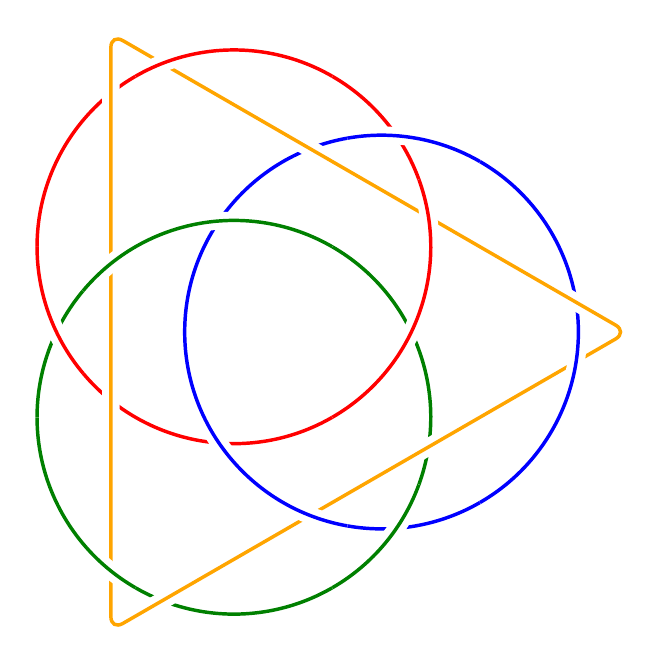}
  \caption{}
  \label{fig:Borrolength4-1}
\end{figure}

\begin{figure}[H]
  \centering
  \includegraphics[width=\ScaleIfNeeded]{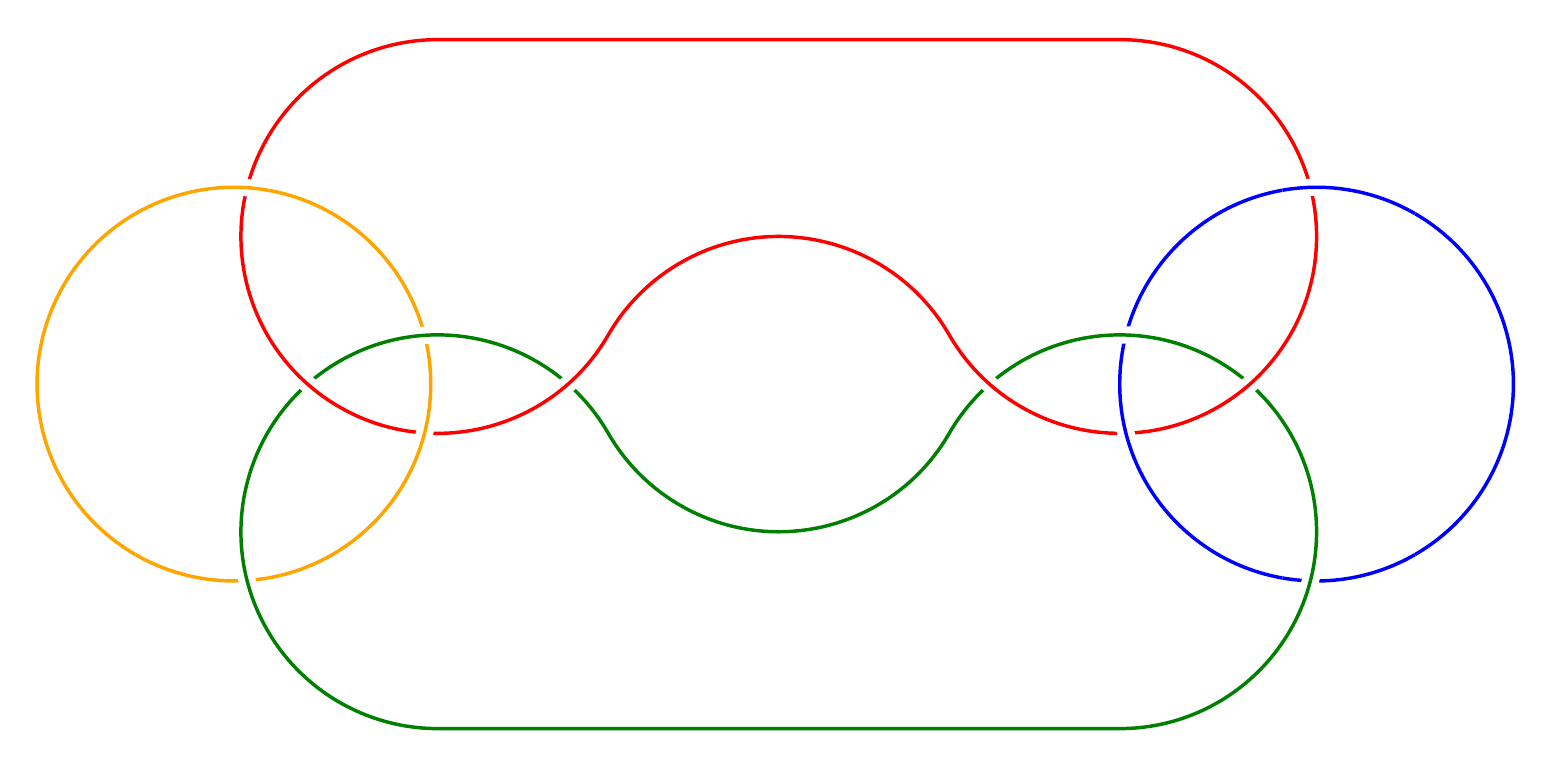}
  \caption{}
  \label{fig:Borrolength4-2}
\end{figure}

\section{A different $2B$-ring}

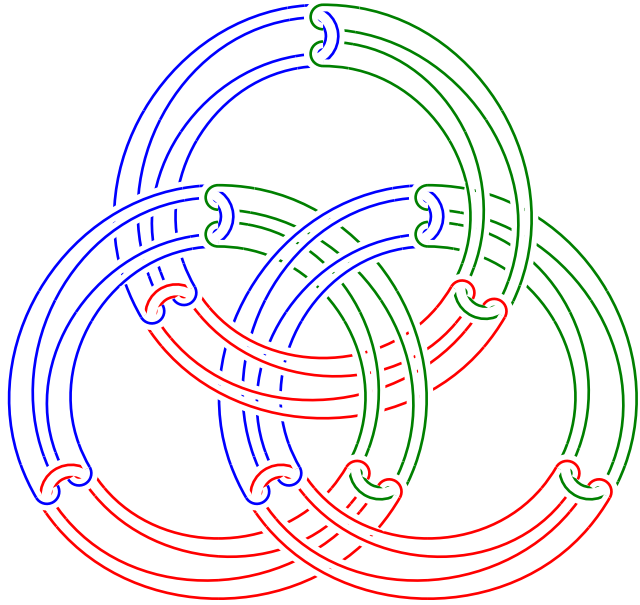
\begin{figure}[H]
  \centering
  \begin{tikzpicture}[scale=0.65]
    \useasboundingbox (-7,-8) rectangle (7,5);
    \pgfmathsetmacro{\brscale}{3}
    \foreach \brl in {1,2,3} {
      \pgfmathsetmacro{\brxshift}{(\brl - 2) * (\brscale + 5*\brstep)
        * cos(60)}
      \pgfmathsetmacro{\bryshift}{-(Mod(\brl,2)*(\brscale + 5*\brstep)
        * sin(60)}
      \begin{scope}[xshift = \brxshift cm, yshift = \bryshift cm]
        \foreach \brk in {1,2,3} {
          \begin{scope}[rotate=\brk * 120 - 60]
            \begin{scope}[rotate=-115,every
              path/.style={knot,double=ring\brk}]
              \draw (0,-\brscale) arc (90:270:2.5*\brstep);
              \draw (0,-\brscale-\brstep) arc (90:270:1.5*\brstep);
              \draw (0,-\brscale) arc (-90:-85:\brscale) coordinate
                (a\brk);
              \draw (0,-\brscale-\brstep) arc
                (-90:-85:\brscale+\brstep) coordinate (b\brk);
              \draw (0,-\brscale-4*\brstep) arc
                (-90:-85:\brscale+3*\brstep) coordinate (c\brk); 
              \draw (0,-\brscale-5*\brstep) arc
                (-90:-85:\brscale+4*\brstep) coordinate (d\brk); 
            \end{scope}

            \begin{pgfonlayer}{back}
              \begin{scope}[every path/.style={knot,double=ring\brk}]
                \draw (0,-\brscale-2*\brstep)
                  arc(-90:-100:\brscale+2*\brstep) coordinate (B\brk);
                \draw (0,-\brscale-3*\brstep)
                  arc(-90:-100:\brscale+3*\brstep) coordinate (C\brk);
              \end{scope}
            \end{pgfonlayer}
            \begin{pgfonlayer}{front}
              \begin{scope}[every path/.style={knot,double=ring\brk}]
                \draw (0,-\brscale) arc (90:-90:\brstep);
                \draw (0,-\brscale) arc (-90:-100:\brscale) coordinate
                  (A\brk);
                \draw (0,-\brscale-3*\brstep) arc (90:-90:\brstep);
                \draw (0,-\brscale-5*\brstep) arc
                  (-90:-100:\brscale+5*\brstep)coordinate (D\brk);
              \end{scope}
            \end{pgfonlayer}
            \begin{scope}
              \path[clip] (0,0) -- (150:\brscale+6*\brstep)
                arc(150:210:\brscale+6*\brstep) -- (0,0);
              \begin{scope}[every
                path/.style={knot,double=ring\brk}]
                \draw (a\brk) to[out=-110,in=170] (A\brk);
                \draw (b\brk) to[out=-110,in=170] (B\brk);
                \draw (c\brk) to[out=-110,in=170] (C\brk);
                \draw (d\brk) to[out=-110,in=170] (D\brk);
              \end{scope}
              \begin{pgfonlayer}{back}
                \begin{scope}
                  \path[clip] (0,0) -- (210:\brscale+6*\brstep)
                    arc(210:270:\brscale+6*\brstep) -- (0,0);
                  \begin{scope}[every
                    path/.style={knot,double=ring\brk}]
                    \draw (a\brk) to[out=-110,in=170] (A\brk);
                    \draw (b\brk) to[out=-110,in=170] (B\brk);
                    \draw (c\brk) to[out=-110,in=170] (C\brk);
                    \draw (d\brk) to[out=-110,in=170] (D\brk);
                  \end{scope}
                \end{scope}
              \end{pgfonlayer}
            \end{scope}
          \end{scope}
        }
      \end{scope}
    }
  \end{tikzpicture}
  \caption{Combining Brunnian and Borromean rings in Figure
    \ref{fig:1} into a second order ring}
  \label{fig:}
\end{figure}

\section{One more level of bending (folding)}
\label{app:bending}

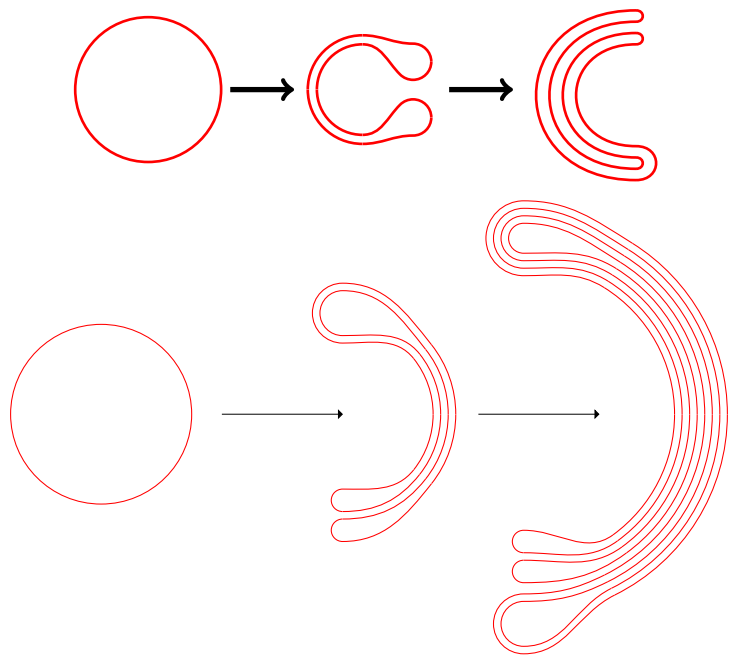
\begin{figure}[H]
  \centering
  \begin{tikzpicture}[every path/.style={knot,double=ring1},scale=0.6]
    \draw (-4.7,0) circle (1.6);
    \colorlet{chain}{ring1}
    \flatbrunnianlink{2}

    \foreach \brk in {1,...,8} {
      \pgfmathsetmacro{\brl}{(\brk > 4 ? -\brk  : 8 - \brk)* \brstep}
      \coordinate (a\brk) at (6,\brl);
    }
    \draw (a1) to[out=0,in=0,looseness=2] (a2)
      (a3) to[out=0,in=0,looseness=2] (a4)
      (a5) to[out=0,in=0,looseness=2] (a8)
      (a6) to[out=0,in=0,looseness=2] (a7)
      (a1) to[out=180,in=180,looseness=2] (a8)
      (a2) to[out=180,in=180,looseness=2] (a7)
      (a3) to[out=180,in=180,looseness=2] (a6)
      (a4) to[out=180,in=180,looseness=2] (a5)
    ;
    \draw[black,double=none,->] (-2.9,0) -- ++(1.4,0);
    \draw[black,double=none,->] (1.9,0) -- ++(1.4,0);
  \end{tikzpicture}
  \\
  \includegraphics[width=\ScaleIfNeeded]{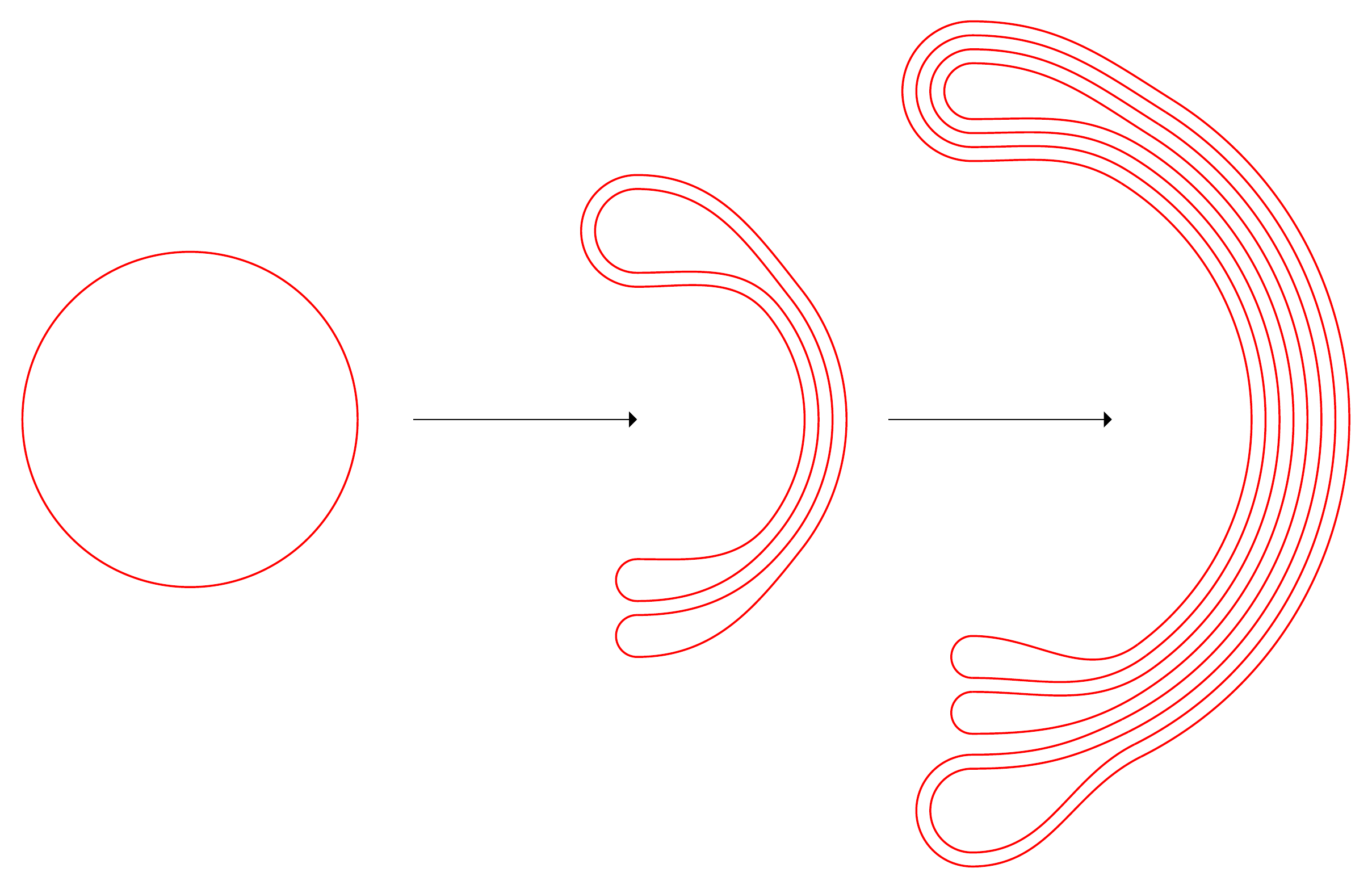}
  \caption{The levels of bending involved --- for construction and
    drawing}
  \label{fig:}
\end{figure}

\newpage

\bibliography{plain}
\nocite*

\end{document}